%% file: TIT_DDBA.tex
\documentclass[journal, draftcls, onecolumn]{IEEEtran}

\usepackage[utf8]{inputenc} 
\usepackage[T1]{fontenc}
\usepackage{url}
\usepackage{ifthen}
\usepackage{cite}
\usepackage[cmex10]{amsmath} 

\usepackage[normalem]{ulem}

\input{./macro_ihwang.tex}

\input{./macro_DDBA.tex}

\begin{document}
\allowdisplaybreaks
\title{Anonymous Heterogeneous Distributed Detection: Optimal Decision Rules, Error Exponents, and the Price of Anonymity}

\author{Wei-Ning Chen and~I-Hsiang~Wang
\thanks{The material in this paper is presented in part at the IEEE International Symposium on Information Theory, Vail, Colorado, USA, June 2018.}
\thanks{W.-N. Chen is with the Graduate Institute of Communication Engineering, 
National Taiwan University, Taipei 10617, Taiwan (email: r05942078@ntu.edu.tw).}
\thanks{I.-H. Wang is with the Department of Electrical Engineering and the Graduate Institute of Communication Engineering, National Taiwan University, Taipei 10617, Taiwan (email: ihwang@ntu.edu.tw).}}


\maketitle

\begin{abstract}
\input{./abstract.tex}
\end{abstract}


\section{Introduction}\label{sec:intro}
\input{./sec_intro.tex}

\section{Problem Formulation and Preliminaries}\label{sec:formulation}
\input{./sec_formulation.tex}

\section{Main Results}\label{sec:results}
\input{./sec_results.tex}

\section{Proof of Theorem~\ref{thm:optimal_test}}\label{sec:proof_optimal_test}
\input{./sec_proof_optimal.tex}
\section{Proof of Theorem~\ref{thm:asymptotics}}\label{sec:proof_asymptotics}
\input{./sec_proof_asymptotics.tex}
\section{A Geometrical Perspective in Chernoff's Regime}\label{sec:chernoff}
\input{./sec_chernoff.tex}
\section{Discussion}\label{sec:discussion}
\input{./sec_discussion.tex}

\section{Conclusion}\label{sec:conclusion}
\input{./sec_conclusion.tex}



\bibliographystyle{IEEEtran}
\bibliography{./DDBA.bib}

\appendices 
\section{Proof of Proposition~\ref{prop:cvx}}\label{sec:proof_cvx}
\input{./appendix_prop.tex}

\section{Proof of Lemma~\ref{lemma:general_sanov}}\label{sec:proof_general_sanov}
\input{./appendix_general_sanov.tex}

\section{Proof of Lemma~\ref{lemma:property_d}}\label{sec:proof_property_d}
\input{./appendix_property_d.tex}


\end{document}

%% file: macro_ihwang.tex
\newcommand{\lp}{\left(}
\newcommand{\rp}{\right)}
\newcommand{\lb}{\left[}
\newcommand{\rb}{\right]}
\newcommand{\lbp}{\left\{}
\newcommand{\rbp}{\right\}}
\newcommand{\lba}{\left\lvert}
\newcommand{\rba}{\right\rvert}

\newcommand{\ul}{\underline}
\newcommand{\ol}{\overline}
\newcommand{\mcal}{\mathcal}

\newcommand{\mbb}{\mathbb}
\newcommand{\msf}{\mathsf}

\newcommand{\ra}{\rightarrow}

\newcommand{\lgra}{\longrightarrow}

\newcommand{\eqDef}{\triangleq}

\newcommand{\E}{\mathbb{E}}

\newcommand{\Ber}{\mathrm{Ber}}

\newcommand{\KLD}[2]{{D}\left( #1\, \middle\Vert #2 \right)}

%% file: macro_DDBA.tex
\usepackage{amsmath,amssymb,mathrsfs}
\usepackage{thm-restate}
\usepackage{bm}
\usepackage{bbm}
\usepackage{amssymb}
\usepackage{graphicx}
\usepackage{color}
\usepackage{xcolor}
\usepackage{subfig}
\usepackage{algorithm}
\usepackage{algorithmic}

\usepackage{eqparbox}

\usepackage{hyperref}
\hypersetup{colorlinks,breaklinks,
            linkcolor=[RGB]{0,0,205},
            citecolor=[RGB]{34,139,34},
            urlcolor=[RGB]{0,0,128}
            }

\usepackage{amsthm}

\theoremstyle{plain}
\newtheorem{lemma}{Lemma}[section]
\newtheorem{theorem}{Theorem}[section]
\newtheorem{proposition}{Proposition}[section]

\newtheorem{definition}{Definition}[section]

\newtheorem{remark}{Remark}[section]

\theoremstyle{remark}
\newtheorem{proofpart}{Part}

\newcommand{\Pnull}{\mbb{P}_{0;\sigma}}
\newcommand{\Palt}{\mbb{P}_{1;\sigma}}


%% file: abstract.tex
We explore the fundamental limits of heterogeneous distributed detection in an anonymous sensor network with $n$ sensors and a single fusion center. The fusion center collects the single observation from each of the $n$ sensors to detect a binary parameter. The sensors are clustered into multiple groups, and different groups follow different distributions under a given hypothesis. The key challenge for the fusion center is the anonymity of sensors -- although it knows the exact number of sensors and the distribution of observations in each group, it does not know which group each sensor belongs to. It is hence natural to consider it as a composite hypothesis testing problem. 
First, we propose an optimal test called \emph{mixture likelihood ratio test}, which is a randomized threshold test based on the ratio of the uniform mixture of all the possible distributions under one hypothesis to that under the other hypothesis. Optimality is shown by first arguing that there exists an optimal test that is \emph{symmetric}, that is, it does not depend on the order of observations across the sensors, and then proving that the mixture likelihood ratio test is optimal among all symmetric tests. 
Second, we focus on the Neyman-Pearson setting and characterize the error exponent of the worst-case type-II error probability as $n$ tends to infinity, assuming the number of sensors in each group is proportional to $n$. 
Finally, we generalize our result to find the collection of all achievable type-I and type-II error exponents, showing that the boundary of the region can be obtained by solving a convex optimization problem.
Our results elucidate the price of anonymity in heterogeneous distributed detection, and can be extended to $M$-ary hypothesis testing with heterogeneous observations generated according to hidden latent variables. The results are also applied to distributed detection under Byzantine attacks, which hints that the conventional approach based on simple hypothesis testing might be too pessimistic. 

%% file: sec_intro.tex
In wireless sensor networks, the cost of identifying individual sensors increases drastically as the number of sensors grows. For distributed detection \cite{Tsitsiklis_90}, when the observations follow identical and independent distributions (i.i.d.) across all sensors, identifying individual sensors is not very important. When the fusion center can fully access the observations, the empirical distribution (types) of the collected observation is a sufficient statistic. When the communication between each sensor and the fusion center is limited, for binary hypothesis testing it is asymptotically optimal to use the same local decision function at all sensors \cite{Tsitsiklis_88}.
Hence, anonymity is not a critical issue for the classical (homogeneous) distributed detection problem. 

However, when the joint distribution of the observations is \emph{heterogeneous}, that is, marginal distributions of observations vary across sensors, sensor anonymity may deteriorate the performance of distributed detection, even for binary hypothesis testing. One such example is distributed detection under Byzantine attack \cite{MaranoMatta_09}, where a fixed number of sensors are compromised by malicious attackers and report fake observations following certain distributions. Even if the fusion center is aware of the number of compromised sensors and the attacking strategy that renders worst-case detection performance (the least favorable distribution as considered in \cite{Huber_65,HuberStrassen_73,VeeravalliBasar_94}), it is more difficult to detect the hidden parameter when the fusion center does not know which sensors are compromised.

In this paper, we aim to quantify the performance loss due to sensor anonymity in heterogeneous distributed detection, with $n$ sensors and a single fusion center. Each sensor (say sensor $i$, $i\in\{1,...,n\}$) has a single random observation $X_i$. The goal of the fusion center is to estimate the hidden parameter $\theta\in\{0,1\}$ (that is, binary hypothesis testing) from the collected observations. The distributions of the observations, however, are \emph{heterogeneous} -- observations at different sensors may follow different sets of distributions. In particular, we assume that these $n$ sensors are clustered into $K$ groups $\{\mcal{I}_1,...,\mcal{I}_K\}$, and group $\mcal{I}_k \subseteq \{1,...,n\}$ comprises $n_k$ sensors, for $k=1,...,K$. 
Under hypothesis $\mcal{H}_\theta$, $\theta\in\{0,1\}$, 
\begin{equation*}
X_i \sim P_{\theta;k},\ \text{for $i\in \mcal{I}_k$.}
\end{equation*}
Moreover, the sensors are \emph{anonymous}, that is, the collected observations at the fusion center are \emph{unordered}. In other words, although the fusion center is fully aware of the \emph{heterogeneity} of it observation, including the set of distributions $\{P_{\theta;k}\mid\theta\in\{0,1\},\ k=1,...,K\}$ and $\{n_k\mid k=1,...,K\}$, it does not know what distribution each individual sensor will follow. 

To address the lack of knowledge about the exact distributions of the observations, we formulate the detection problem as a \emph{composite hypothesis testing} problem, where the vector observation of length $n$ follows a product distribution within a finite class of $n$-letter product distributions under a given parameter $\theta$. The class consists of $\binom{n}{n_1,...,n_K}$ possible product distributions, each of which follows one of the $\binom{n}{n_1,...,n_K}$ possible partitions of the sensors. The fusion center takes all the possible partitions into consideration when detecting the hidden parameter. 
We mainly focus on a Neyman-Pearson setting, where the goal is to minimize the worst-case type-II error probability such that the worst-case type-I error probability is not larger than a constant. Towards the end of this paper, we also extend our results to a Bayesian setting, where a binary prior distribution is laid on $\mcal{H}_0$ and $\mcal{H}_1$. 

Our main contribution comprises three parts. First, we develop an optimal test, termed \emph{mixture likelihood ratio test} (MLRT), for the anonymous heterogeneous distributed detection problem. MLRT is a randomized threshold test based on the ratio of the uniform mixture of all the possible distributions under hypothesis $\mcal{H}_1$ to the uniform mixture of those under $\mcal{H}_0$. To prove the optimality, we first argue that there exists an optimal test that is \emph{symmetric}, that is, it does not depend on the order of observations across the sensors, and thus we only need to consider tests which depend on the histogram of observations. In other words, the histogram of observations contains sufficient information for optimal detection. Moreover, all possible distributions over the space of observations $\mcal{X}^n$ under $\mcal{H}_0$ (or $\mcal{H}_1$) turn out to be the same one over the space of its histogram, so if we test the hypothesis according to the histogram, the original composite hypothesis testing problem boils down to a simple hypothesis testing problem. 
The one-to-one correspondence between symmetric tests and tests defined on the histogram is the key to derive optimal test. This result extends to $M$-ary hypothesis testing with heterogeneous observations generated according to hidden latent variables, each of which is associated to a observation, but the decision maker only knows the histogram of the latent variables.


Second, for the case that the alphabet $\mcal{X}$ is a finite set, we characterize the error exponent of the minimum worst-case type-II error probability as $n\ra\infty$ with the ratios $\frac{n_k}{n} \ra \alpha_k$ $\forall\,k=1,...,K$. The optimal error exponent turns out to be the minimization of a linear combination of Kullback-Leibler divergences (KL divergences) with the $k$-th term being $\KLD{U_k}{P_{1;k}}$ and $\alpha_k$ being the coefficient, for $k=1,...,K$. The minimization is over all possible distributions $U_1,...,U_K$ such that $\sum_{k=1}^{K}\alpha_kU_k = \sum_{k=1}^{K}\alpha_kP_{0;k}$. In a simple hypothesis testing problem with i.i.d. observations, a standard approach to derive the type-II error exponent is invoking a strong converse lemma (see, for example, Chapter~12 in \cite{PolWu17}) to relate the type-I and type-II error probability of an optimal test, and then applying the large deviation toolkit on the optimal test to single-letterize and find the exponent.
In contrast, in our problem, neither can the mixture distributions in the optimal test be decomposed into a product form, nor can the acceptance region be bounded by a large deviation event, making this approach fail to characterize the error exponent. To circumvent the difficulties, we turn to the method of types and use bounds on types (empirical distributions) for single-letterization.  

For achievability, instead of the optimal MLRT which is difficult to single-letterize, we employ a simpler test that resemble Hoeffding's test \cite{Hoeffding_65}. For the converse, we use an argument based on the method of types. We propose a generalized divergence $D_{\alpha_1,...,\alpha_K}(P_1,...,P_K ; Q_1,...,Q_K)$ from a group of distributions $\{Q_1,...,Q_K\}$ to another group of distributions $\{P_1,...,P_K\}$, which plays a similar role as KL divergence in simple hypothesis testing problems. The key to the characterization of the optimal error exponent is to prove a generalized Sanov Theorem for the composite setting we considered. Based on the characterized error exponent, given the number of bits that a sensor can send to the fusion center, one can also formulate an optimization problem to find the best local decision functions, as in the homogeneous case \cite{Tsitsiklis_88}.



Finally, we extend our results from the Neyman-Pearson setting to a Bayesian setting, minimizing the average probability of error (that is, combining type-I and type-II error). It can be shown that the optimal test is computationally infeasible, since it involves summation over all possible permutations. To overcome the complexity issue, we propose an asymptotically optimal test based on information geometry, which achieves the same error exponent of the average probability of error. We also study the exponent region $\mcal{R}$, the collection of all pairs of achievable type-I and type-II error exponents. In particular, we propose a way to parametrize the contour of $\mcal{R}$ based on information projection. However, the closed-form expression of $\mcal{R}$ involves an explicit solution of a convex optimization problem, which remains unsettled.

As a by-product, we apply our results for $K=2$ to the distributed detection problem under Byzantine attack and further obtain bounds on the worst-case type-II error exponent.
Compared with the worst-case exponent in an alternative Bayesian formulation \cite{MaranoMatta_09} where the observation of sensors are assumed to be i.i.d. according to a mixture distribution, it is shown that the worst-case exponent in the composite testing formulation is strictly larger. This hints that the conventional approach taken in \cite{MaranoMatta_09} might be too pessimistic.  

\subsection*{Related Works}

Decentralized detection is a classical topic, and attracts extensive attention in recent years due to its application in wireless sensor networks. See, for example, \cite{Tsitsiklis_88,Tsitsiklis_90,TenSan81,VeeravalliBasar_94}. Most works in decentralized detection are focused on finding optimal local decision function in both Neyman-Pearson and Bayesian regime. Under some assumptios on the distribution of a given hypothesis, optimal design criteria of local decision function and the decision rule at the fusion center are given. 
Unlike the anonymous setting considered in our work, the above-mentioned classical works assume fusion centers, as well as the local sensors, have perfect knowledge about the joint distribution, and hence the decision rules are designed according to it. This is termed an ``informed" setting in our paper and is used as a baseline to compare with and see the price of anonymity.
On the other hand, in our setting, the fusion center collects observations without knowing the exact index of each one, and thus the problem is formulated into a composite hypothesis testing problem.

Composite hypothesis testing is a long-standing problem in statistics, and is notoriously difficult to find an optimal test. In general, the uniform most powerful (UMP) test does not exist, see, for example, Section~8.3 in \cite{GeoBer02}. Even if we relax the performance evaluation to the \emph{minimax} regime, the general form of the optimal test is still unknown, except for some special case. For example, \cite{HuberStrassen_73} considered the case that the composite hypothesis class $\mcal{H}_\theta$ is formed by all $\epsilon$-contaminated distributions of $P_\theta$, that is, $\lbp (1-\epsilon)P_\theta+ \epsilon Q \mid \forall\, \text{possible distributions $Q$}\rbp$. Under this structure, Huber showed that a censored version of likelihood ratio test is optimal in the minimax regime. Other works such as \cite{Hoeffding_65,ZeitouniGutman_91} followed the idea of Hoeffding's test \cite{Hoeffding_65} and proposed an universal asymptotically optimal test when the null hypothesis is simple. Meanwhile, in our setting, neither the parameter space of the considered distributions is continuous, nor the null hypothesis is simple, making their approaches hard to extend.  Another common test for composite hypothesis testing is the \emph{generalized likelihood ratio test} (GLRT). The optimality of GLRT is guaranteed under some circumstances, see, for example, \cite{ZeitouniZiv_92}. However, the results in \cite{ZeitouniZiv_92} hold only for simple null and composite alternative. 
In contrast, our result indicates that GLRT is not optimal in our setting.

The concept of Byzantine attack can be traced back to \cite{LamShoPea82} (known as the ``Byzantine Generals Problem''), in which reliability of a computer system with malfunctioned components is studied. After that, Byzantine model is developed and generalized by several research areas, especially in communication security. For example, the distributed detection with Byzantine attack is studied under the Neyman-Pearson formulation in \cite{MaranoMatta_09} and under the Bayesian setting in \cite{KaiHanBra15}. In their settings, each sensor is assumed to be compromised with probability $\alpha$, so the observation turns out to be drawn identically and independently from an mixture distribution, making the hypothesis testing problem simple, and thus Neyman-Pearson lemma can be applied. In contrast, in our work we assume the number of Byzantine sensors is fixed and is $\alpha n$, where $n$ is the total number of sensors, and thus the problem falls into a composite hypothesis testing instead of the mixture setting.

%

This work is presented in part at ISIT 2018. In the conference version \cite{CheCheWan18}, upper and lower bounds on the type-II error exponent were given, where the lower bound (achievability) is based on an modified version of Hoeffding's test, and the upper bound (converse) is derived by relaxing the original problem into a simple hypothesis testing. In this journal version, we show that the achievability bound in the conference version is indeed tight, closing the gap between the upper and lower bounds.

The rest of this paper is organized as follows. In Section~\ref{sec:formulation}, we formulate the composite hypothesis testing problem for anonymous heterogeneous distributed detection and provide some background. In Section~\ref{sec:results}, the main results are provided, where the proofs are delegated to Section~\ref{sec:proof_optimal_test} and \ref{sec:proof_asymptotics}. In Section~\ref{sec:chernoff}, we generalize the results to the Bayesian setting, and in Section~\ref{sec:discussion}, we briefly discuss the case when $\mcal{X}$ is not finite, and the case when partial information about the group assignment is available at the fusion center.
Finally, we conclude the paper with some further directions and open questions in Section~\ref{sec:conclusion}.

%% file: sec_formulation.tex
\subsection{Problem Setup}
Following the description of the setting in Section~\ref{sec:intro}, let us formulate the composite hypothesis testing problem. Let $\sigma(i)$ denote the label of the group that sensor $i$ belongs to. This labeling $\sigma(\cdot)$, however, is not revealed to the fusion center. Hence, the fusion center needs to consider all $\binom{n}{n_1,...,n_K}$ possible $\sigma: \{1,...,n\}\ra\{1,...,K\}$ satisfying 
\begin{equation}\label{eq:condition}
\lba\{i\mid \sigma(i)=k\}\rba = n_k,\ \forall\,k=1,...,K,
\end{equation}
and decides whether the hidden $\theta$ is $0$ or $1$. 
For notational convenience, let $\bm{\nu}$ denote the vector $[n_1\ ...\ n_K]^\intercal$, and let $\mcal{S}_{n,\bm{\nu}}$ denote the collection of all labelings satisfying \eqref{eq:condition}.

Hence, the fusion center is faced with the following \emph{composite} hypothesis testing problem, where the goal is to infer the parameter $\theta$:
\begin{equation*}\textstyle
\mcal{H}_\theta: X^n\sim \mbb{P}_{\theta;\sigma} \eqDef \prod_{i=1}^n P_{\theta;\sigma(i)},\ \text{for some $\sigma\in\mcal{S}_{n,\bm{\nu}}$}.
\end{equation*}
As mentioned in Section~\ref{sec:intro}, throughput the paper we consider binary hypothesis testing, that is, $\theta \in \{0,1\}$.

Let each single observation take values from some measurable space $(\mcal{X}, \mcal{F})$, where $\mcal{F}$ is a $\sigma$-algebra on $\mcal{X}$. Hence $P_{\theta;k}\in\mcal{P_X}$ for all $\theta\in\{0,1\}$ and $k\in\{1,...,K\}$, where $\mcal{P_X}$ denotes the collection of all possible distributions over $(\mcal{X}, \mcal{F})$. 
The vector observation $x^n$ is defined on the space $(\mcal{X}^n, \mcal{F}^{\otimes n})$, where $\mcal{F}^{\otimes n}$ is the \emph{tensor product} $\sigma$-algebra of $\mcal{F}$, that is, the smallest $\sigma$-algebra contains the following collection of events:
$$ \lbp \mcal{E}_1\times\mcal{E}_2\times \cdots \times \mcal{E}_n \mid \mcal{E}_i \in \mcal{F} \rbp.  $$

A (randomized) test is a measurable function $\phi: \lp \mcal{X}^n, \mcal{F}^{\otimes n}\rp \ra \lp [0,1], \mathfrak{B}\rp$, where $\mathfrak{B}$ denotes the Borel $\sigma$-field on $\mbb{R}$.
The worst-case type-I and type-II error probabilities of a decision rule $\phi$ are defined as 
\begin{align*}
\msf{P_{F}}^{(n)}(\phi) &\eqDef \max_{\sigma\in\mcal{S}_{n,\bm{\nu}}} \E_{\mbb{P}_{0;\sigma}}\lb \phi(X^n) \rb \quad \text{(Type I)}\\
\msf{P_{M}}^{(n)}(\phi) &\eqDef \max_{\sigma\in\mcal{S}_{n,\bm{\nu}}} \E_{\mbb{P}_{1;\sigma}}\lb 1-\phi(X^n) \rb \quad \text{(Type II)}.
\end{align*}
Our focus is on the Neyman-Pearson setting: find a decision rule $\phi$ satisfying $\msf{P_{F}}^{(n)}(\phi) \le \epsilon$ such that $\msf{P_{M}}^{(n)}(\phi)$ is minimized. Let $\beta^{(n)}(\epsilon,\bm{\nu})$ denote the minimum type-II error probability.

For the asymptotic regime, we assume that the ratio $\frac{n_k}{n}\ra \alpha_k$ as $n\ra\infty$ for all $k=1,...,K$, and $\sum_{k=1}^K\alpha_k = 1$.
We aim to explore if $\beta^{(n)}(\epsilon,\bm{\nu})$ decays exponentially fast as $n\ra\infty$, and characterize the corresponding error exponent. For notational convenience, we define upper and lower bounds on the exponent:
\begin{align*}
\ol{E}^*(\epsilon,\bm{\alpha}) 
&\textstyle
\eqDef \limsup_{n\ra\infty} \lbp-\frac{1}{n}\log_2\beta^{(n)}(\epsilon,\bm{\nu})\rbp,\\
\ul{E}^*(\epsilon,\bm{\alpha}) 
&\textstyle
\eqDef \liminf_{n\ra\infty} \lbp-\frac{1}{n}\log_2\beta^{(n)}(\epsilon,\bm{\nu})\rbp,
\end{align*}
where in taking the limits, we assume that $\lim_{n\ra\infty}\frac{n_k}{n} = \alpha_k$, for all $k=1,...,K$.
If the upper and lower bound match, we simply denote it as $E^*(\epsilon,\bm{\alpha})$.

\begin{remark}
The original distributed detection problem \cite{Tsitsiklis_90,Tsitsiklis_88,VeeravalliBasar_94} involves local decision functions at the sensors to address the limited communication between each sensor and the fusion center. In order to focus on the impact of anonymity, we first absorb them into the distributions $\{P_{\theta;k}:k=1,...,K\}$ because they are symbol-by-symbol maps. Later, we will discuss how to find the best local decision functions according to the characterized error exponent.

\end{remark}

\subsection{Notations}
Let us introduce notations that will be used throughout this paper.
\begin{itemize}

	\item $n$ denotes the total number of observations, and $K$ denotes the number of groups of sensors.
	\item  $\bm{\nu} \eqDef [n_1\ ...\ n_K]^\intercal$ denotes the number of sensors in the $K$ groups. That is, $n_k \geq 0$, $n_k\in\mbb{Z}$, and $\sum_{k=1}^K n_k = n$.
	\item $\bm{\alpha} \eqDef [\alpha_1\ ...\ \alpha_K]^\intercal$ denotes the fraction of each group of sensors in all sensors in the asymptotic regime. That is, $\alpha_k \geq 0$, and $\sum_{k=1}^K \alpha_k = 1$. 

	\item $\sigma : \{1,...,n\} \ra \{1,...,K\}$ is the labeling function which assigns the index of each sensor to a group. We also denote the collection of indices of sensors in group $k$ as 
	\begin{equation}\label{eq:card_constraint}
		\mcal{I}_k = \sigma^{-1}(k) \eqDef \lbp i \mid \, \sigma(i) = k \rbp.
	\end{equation} 

	\item Let $\mcal{S}_{n,\bm{\nu}}$ be the collection of all $\sigma$ satisfying \eqref{eq:card_constraint}. We also use $\mcal{S}_n$ to denote the collection of length-$n$ permutations:
	$$ \mcal{S}_n \eqDef \lbp \tau : \{1,2,...,n\} \overset{1-1}{\ra}\{1,2,...,n\}\rbp.$$
	Note that the cardinalities of the two sets are 
	$$\lba\mcal{S}_{n,\bm{\nu}}\rba = {n\choose {n_1, n_2,...,n_K}}, \, \lba \mcal{S}_n \rba = n!.$$
	
	\item We usually write $\bm{P}_\theta$ as the vector of $\{P_{\theta;k}\}$:
	\[
	\bm{P}_\theta \eqDef
	\begin{bmatrix}
    	P_{\theta;1}\\
    	P_{\theta;2}\\
    	\vdots\\
    	P_{\theta;K}\\
	\end{bmatrix}.
	\]
\end{itemize}

\subsection{Method of Types}
For a sequence $x^n \in \mcal{X}^n$, where $\mcal{X} = \{a_1,a_2,...,a_d\}$, its type (empirical distribution) is defined as 
	$$ \Pi_{x^n} = \lb \pi(a_1|x^n), \pi(a_2|x^n),...,\pi(a_d|x^n) \rb, $$ where $\pi(a_i|x^n)$ is the frequency of $a_i$ in the sequence $x^n$, that is, 
	$$ \pi(a_i|x^n) = \frac{1}{n}\sum_{j=1}^{n}\mathbbm{1}_{\lbp x_j=a_i\rbp}.$$
	For a given length $n$, we use $\mcal{P}_n$ to denote the collection of possible types of length-$n$ sequences. In other words, 
	$$ \mcal{P}_n \eqDef \lbp\lb \frac{i_1}{n}, \frac{i_2}{n},..., \frac{i_d}{n}\rb \bigg\vert\, \forall i_1,...,i_d \in \mbb{N}\cup \{0\}, i_1+i_2+\cdots+i_d = n \rbp. $$
	Let $U\in \mcal{P}_n$ be an $n$-type. The type class $T_n(U)$ is the set of all length-$n$ sequences with type $U$, 
	$$ T_n(U) \eqDef \lbp x^n \in \mcal{X}^n \mid \Pi_{x^n} = U \rbp.$$ 

Let us introduce some useful lemmas about type.
\begin{lemma}[Cardinality Bound of $\mcal{P}_n$]\label{lemma:Pn_bound}
$$ \lba \mcal{P}_n \rba \leq (n+1)^{\lba \mcal{X} \rba}.$$
In words, $\lba\mcal{P}_n\rba$ grows polynomial in $n$.
\end{lemma}
\begin{lemma}[Probability of Type Class]\label{lemma:type}
	 Let $P\in \mcal{P}_n, Q \in \mcal{P_X}$. Then 
	$$ \frac{1}{(n+1)^{\lba\mcal{X}\rba}}2^{-n\KLD{Q}{P}} \leq Q^{\otimes n}(T_n(P))\leq 2^{-n\KLD{Q}{P}}. $$
\end{lemma}

For finite $\mcal{X}$, $\mcal{P_X}$ can be viewed as a subspace in $\mbb{R}^d$ endowed with Euclidean metric and standard topology. The following theorem, developed by Sanov, depicts the probability of a large deviation event.
\begin{lemma}[Sanov's Theorem]\label{lemma:sanov}
Let $\Gamma \subseteq \mcal{P_X}$. Then we have 
\begin{equation}\label{eq:sanov}
 -\inf_{T\in\text{int }\Gamma} \KLD{T}{Q} \leq \liminf_{n\ra\infty} \frac{1}{n}\log Q\lbp x^n: \Pi_{x^n} \in \Gamma \rbp \leq \limsup_{n\ra\infty} \frac{1}{n}\log Q\lbp x^n: \Pi_{x^n} \in \Gamma \rbp \leq -\inf_{T\in\text{cl }\Gamma} \KLD{T}{Q},
 \end{equation}
where $\text{int } \Gamma$ and and $\text{cl } \Gamma$ respectively denote the interior and the closure of $\Gamma$, with respect to the standard topology on $\mbb{R}^d$.
In particular, if the infimum on the right-hand side is equal to the infimum on the left-hand side in \eqref{eq:sanov}, we have 
$$ \lim_{n\ra\infty} \frac{1}{n}\log Q\lbp x^n: \Pi_{x^n} \in \Gamma \rbp = -\inf_{T\in\Gamma} \KLD{T}{Q}. $$
\end{lemma}

Proofs of the lemmas mentioned above can be found in standard information theory textbooks, Chapter~11 in \cite{CovTho06} for example. Alternatively, a more rigorous proof of Sanov's theorem Lemma~\ref{lemma:sanov} can be found in \cite{Csi06}.

%% file: sec_results.tex
As mentioned in Section~\ref{sec:formulation}, the observations come from the measurable space $(\mcal{X}^n, \mcal{F}^{\otimes n})$. Throughout the rest of the paper, we assume that $\mcal{X}$ is a totally ordered set, and  $\mcal{F}^{\otimes n}$ satisfies the following two assumptions:
\begin{enumerate}
	\item $\mcal{F}^{\otimes n}$ contains the following set:
	\begin{equation}\label{eq:type_space}
		\tilde{\mcal{X}}^n \eqDef \lbp (x_1, x_2,...,x_n) \mid x_1 \geq x_2 \geq ... \geq x_n \rbp. 
	\end{equation}
	\item $\mcal{F}^{\otimes n}$ is closed under permutation. That is, if $\mcal{A} \in \mcal{F}^{\otimes n}$, for any length-$n$ permutation $\tau: \{1,...,n\} \ra \{1,...,n\}$, 
	\begin{equation} \label{eq:tau_shift}
	\mcal{A}_\tau \eqDef \lbp \lp x_{\pi(1)}, ..., x_{\pi(n)}\rp \mid (x_1,...,x_n) \in \mcal{A}  \rbp \in \mcal{F}.
	\end{equation}
\end{enumerate}
\begin{remark}
	We assume that $\mcal{X}$ is a totally ordered set in order to set the condition such that $\tilde{\mcal{X}}$ is measurable. The purpose to require $\tilde{\mcal{X}}$ to be measurable is to preserve the measurability of the ordering map $\Pi(\cdot)$, as later defined in Definition~\ref{def:Pi}. In general, if $\mcal{X}$ is not totally ordered, we can still require the collection of representatives in the equivalent classes induced by $\Pi^{-1}$ to be measurable. However, the regularity assumptions on $\mcal{F}^{\otimes}$ need to be carefully concerned in that case.
\end{remark}
\begin{remark}
The second assumption always holds for tensor $\sigma$-fields. The first assumption typically holds too. For example, if $\mcal{X}$ is finite, we can simply choose $\mcal{F}$ as the power set $ 2^{\mcal{X}}$, and if $\mcal{X} \subseteq \mbb{R}$, we can choose $\mcal{F}$ as the Borel $\sigma$-field. 
In particular, for $\mcal{X}$ being a finite set, it is straightforward to define a total order over it, and hence it is a totally ordered set. Moreover, the above two assumptions are automatically satisfied. 
\end{remark}

\subsection{Main Contributions}
Our first contribution is the characterization of the optimal test:
\begin{theorem}[Optimal Test]\label{thm:optimal_test}
	Define the \emph{mixture likelihood  ratio} $\ell (x^n) $:
	\begin{equation} \label{eq:mlr}
	\ell(x^n) \eqDef \frac{\sum_{\sigma\in\mcal{S}_{n,\bm{\nu}}}\mbb{P}_{1;\sigma} (x^n)}{\sum_{\sigma\in\mcal{S}_{n,\bm{\nu}}}\mbb{P}_{0;\sigma} (x^n)}.
	\end{equation}
	Suppose $\mcal{F}^{\otimes n}$ satisfies the two assumptions \eqref{eq:type_space}, \eqref{eq:tau_shift}. Then an optimal tests $\phi^*(x^n)$ takes the following form:
	\begin{equation}\label{eq:optimal_test}
	\phi^*(x^n) = 
	\begin{cases}
	 1, &\text{ if } \ell(x^n) > \tau \\
	 \gamma, &\text{ if } \ell(x^n) = \tau \\
	 0, &\text{ if } \ell(x^n) < \tau.
	\end{cases}
	\end{equation}
	That is, for any test $\phi$, we have 
	$$  \mathsf{P_F}(\phi) \leq \mathsf{P_F}(\phi^*) \Rightarrow \mathsf{P_M}(\phi) \geq \mathsf{P_M}(\phi^*).$$
\end{theorem}
\begin{remark}
	We see that the optimal test, MLRT, is the likelihood ratio test between two uniform mixture distributions
	$$ \frac{1}{\lba\mcal{S}_{n,\bm{\nu}}\rba}\sum_{\sigma\in\mcal{S}_{n,\bm{\nu}}}\mbb{P}_{\theta;\sigma}, 
	\, \theta \in \{0,1\}.$$
	Interestingly, the optimality of MLRT indicates that the widely used decision rule, generalized likelihood ratio test (GLRT), which is defined as the randomized thresholded test according to the following likelihood ratio
	$$ \ell_{\text{GLRT}}(x^n)\eqDef\frac{\sup_{\sigma\in\mcal{S}_{n,\bm{\nu}}}\mbb{P}_{1;\sigma} (x^n)}{\sup_{\sigma\in\mcal{S}_{n,\bm{\nu}}}\mbb{P}_{0;\sigma} (x^n)}, $$
	is strictly sub-optimal in the anonymous hypothesis testing problem.
\end{remark}
\begin{IEEEproof}[Sketch of proof]
	The proof consists of two steps. In the first step, we introduce \emph{symmetric tests} (as later defined in Definition~\ref{def:sym_test}), which do not depend on the order of the observations. Then, we show that among all symmetric tests, \eqref{eq:optimal_test} is optimal. The key is to reduce the original composite hypothesis testing problem into a simple one through the ordering map $\Pi(x^n)$ in Definition~\ref{def:Pi}, and then apply Neyman-Pearson lemma.

	In the second step, we prove that for any test $\psi$, one can always \emph{symmetrize} it and construct a symmetric one $\phi$ which is as good as $\psi$, so \eqref{eq:optimal_test} is optimal among all tests. However, $\psi$ is constructed by assigning values on each equivalence classes introduced by the ordering map $\Pi(\cdot)$, so the measurability of $\psi$ need to be carefully examined. For the detailed proof, please refer to Section~\ref{sec:proof_optimal_test}.
\end{IEEEproof}

Our second result specifies the exponent of type-II error in Neyman-Pearson formulation, which does not depend on the type-I error probability $\epsilon$:
\begin{theorem}[Asymptotic Behavior]\label{thm:asymptotics}
	Let us consider the case $|\mcal{X}| < \infty$, 
	The exponent of type-II error probability is characterized as follows.
	\begin{equation}\label{eq:lowerbd}
	\begin{split}
	E^*(\epsilon,\bm{\alpha}) = &\min\limits_{\bm{U} \in (\mcal{P_X})^K
	}\, \sum\nolimits_{k=1}^K\alpha_k\KLD{U_k}{P_{1;k}}\\
	& \begin{array}{ll}
	\text{subject to} &\bm{\alpha}^\intercal\bm{U}=\bm{\alpha}^\intercal\bm{P}_0.
	\end{array}
	\end{split}
	\end{equation}
\end{theorem}
\begin{remark}\label{rmk:asymptotic}
	A standard way to derive the exponent of type-II error probability is to identify the acceptance region (of $\mcal{H}_0$) of the optimal test \eqref{eq:optimal_test} as an large-deviation event under $\mcal{H}_1$, and further apply a strong converse lemma to obtain a bound. However, notice that the mixture measure, $\sum_{\sigma} \mbb{P}_{\theta;\sigma}, \, \theta\in\{0,1\}$, cannot be factorized into a product form, which makes it hard to single-letterize. Instead, if we add an additional assumption that $\mcal{X}$ is finite, then we can utilize method of types, such as Sanov's theorem, to circumvent the difficulties.  
\end{remark}
\begin{IEEEproof}[Sketch of proof]
	For the achievability part, we propose a sub-optimal test based on Hoeffding's result \cite{Hoeffding_65}, in which we accept observations $x^n$ satisfying $\KLD{\Pi_{x^n}}{M_0(\bm{\alpha})}\leq \epsilon$ for some threshold $\epsilon$. We apply tools in method of types to bound the type-I and type-II error probabilities, showing that \eqref{eq:lowerbd} is achievable.

	For the converse part, given an arbitrary test, we define its acceptance region as $\mcal{A}$ (if the given test is randomized, we can round the test by 1/2 and make it determinstic, that is, we accept $\mcal{H}_1$ if $\phi(x^n)>1/2$) and consider another high-probability set $\mcal{B}$. We analyze the probability of $\mbb{P}_{1;\sigma}\lbp\mcal{A}\cap\mcal{B}\rbp$, and show that the exponent cannot be greater than \eqref{eq:lowerbd}, which concludes the converse part. For the detailed proof, please refer to Section~\ref{sec:proof_asymptotics}. 
\end{IEEEproof}

Finally, we give a structural result of the error exponent.
\begin{proposition}\label{prop:cvx}
	For the case $|\mcal{X}| < \infty$, the type-II error exponent $E^*(\epsilon,\bm{\alpha})$ as characterized in Theorem~\ref{thm:asymptotics} only depends on $\bm{\alpha}$. Moreover, it is a convex function of $\bm{\alpha}$.
\end{proposition}
\begin{IEEEproof}
See Appendix~\ref{sec:proof_cvx}.
\end{IEEEproof}

\subsection{Numerical Evaluations}

To quantify the price of anonymity, note that when the sensors are not anonymous (termed the ``informed'' setting), it becomes a simple hypothesis testing problem, and the error exponent of the type-II probability of error in the Neyman-Pearson setting is straightforward to derive:
\begin{equation*}\textstyle
E^*_{\text{Informed}}(\epsilon,\bm{\alpha}) = \sum_{k=1}^K \alpha_k \KLD{P_{0;k}}{P_{1;k}}.
\end{equation*}

\begin{figure}[htbp]
	\centering
	\subfloat{{\includegraphics[width=0.4\linewidth]{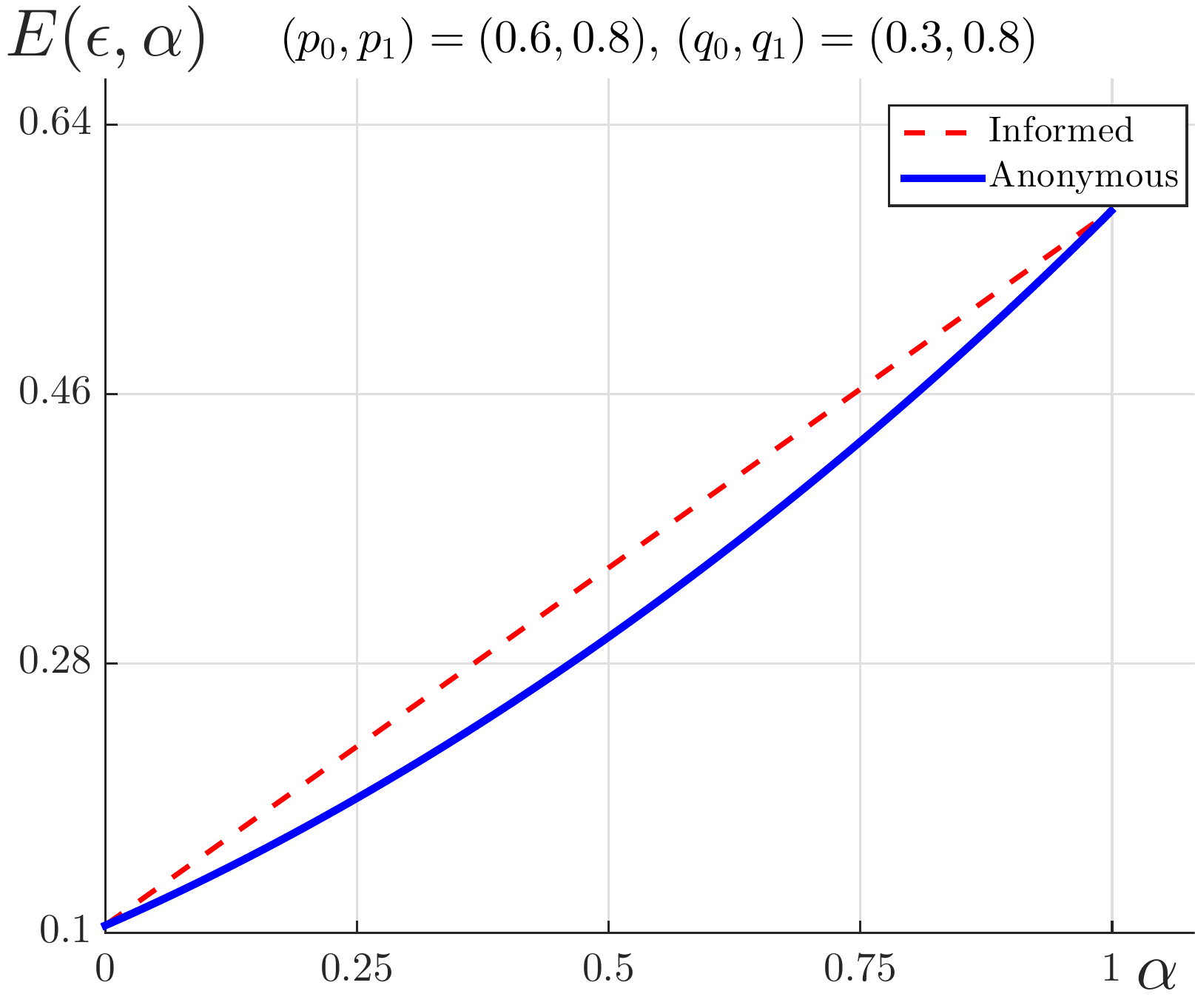} }} %
	\subfloat{{\includegraphics[width=0.4\linewidth]{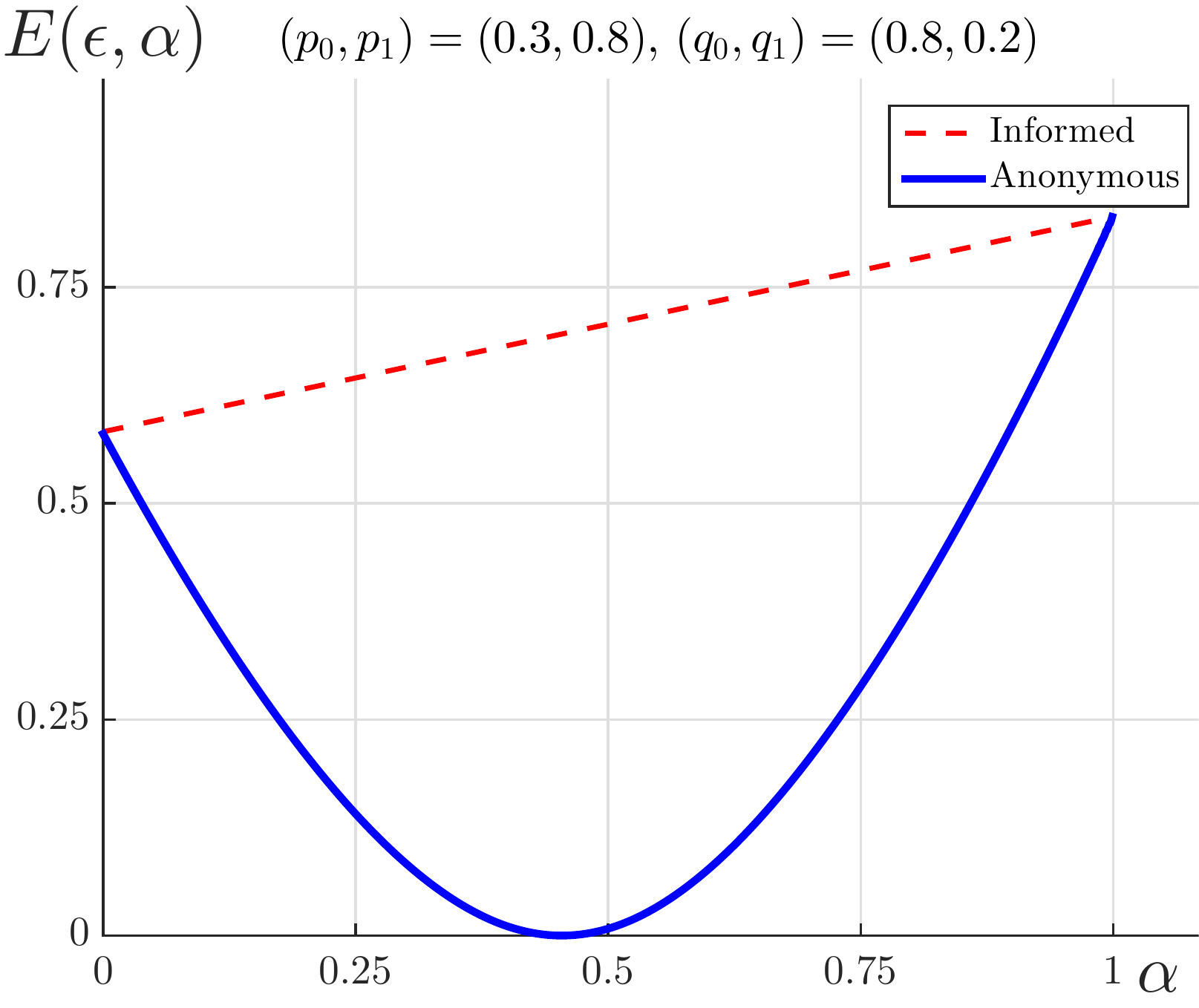} }}%
	\caption{Price of anonymity }%
	\label{fig:anony}
\end{figure}

For ease of illustration, in the following we restrict to the special case of binary alphabet, that is, $|\mcal{X}|=2$, and $K=2$ groups. Let $P_{\theta;1}=\Ber(p_\theta)$ and $P_{\theta;2}=\Ber(q_\theta)$, for $\theta=0,1$, where $\Ber(p)$ is the Bernoulli distribution with parameter $p$. Since there are only two groups, we set $\bm{\alpha} \equiv \begin{bmatrix} 1-\alpha & \alpha\end{bmatrix}^\intercal$. Numerical examples are given in Figure~\ref{fig:anony} to illustrate the price of anonymity versus the mixing parameter $\alpha$. In general, anonymity may cause significant performance loss. In certain regimes, the type-II error exponent can even be pushed to zero. 

\subsection{Distributed Detection with Byzantine Attacks}

Let us apply the results to distributed detection under Byzantine attacks, where the sensors are partitioned into two groups. One group consists of $n(1-\alpha)$ \emph{honest} sensors reporting true i.i.d. observations, while the other consists of $n\alpha$ \emph{Byzantine} sensors reporting fake i.i.d. observations. Here we again neglect the local decision function and assume that each sensor can report its observation to the fusion center. The true observations follow $P_{\theta}$ i.i.d. across honest sensors, while the compromised ones follow $Q_{\theta}$ i.i.d. across Byzantine sensors, for $\theta = 0,1$. In general, $Q_\theta$ is unknown to the fusion center, but in terms of error exponent, one can find the least favorable pair $Q_0,Q_1$ which minimize the error exponent. Hence, our results can be applied here and arrive the worst-case type-II error exponent as follows:
\begin{equation}\label{eq:byzantine_ach}
	\begin{split}
	&\min\limits_{\substack{Q_0,Q_1,U,V\in\mcal{P_X}}}\, (1-\alpha)\KLD{U}{P_1}+\alpha\KLD{V}{Q_1} \\
	&\text{subject to } (1-\alpha)U+\alpha V = (1-\alpha)P_0+\alpha Q_0.
	\end{split}
\end{equation}

In \cite{MaranoMatta_09}, it assumes that each sensor can be compromised with probability $\alpha$, and hence it becomes a homogeneous distributed detection problem, where the observation of each sensor follows a mixture distribution $(1-\alpha)P_{\theta} + \alpha Q_{\theta}$ under hypothesis $\theta$, i.i.d. across all sensors. The worst-case exponent of type-II error probability, as derived in \cite{MaranoMatta_09}, is hence
\begin{equation}\label{eq:byzantine_iid}
\min_{Q_0,Q_1\in\mcal{P_X}} \KLD{(1-\alpha)P_{0} + \alpha Q_{0}}{(1-\alpha)P_{1} + \alpha Q_{1}}.
\end{equation} 

\begin{figure}[htbp]
	\centering
\includegraphics[width=0.6\linewidth]{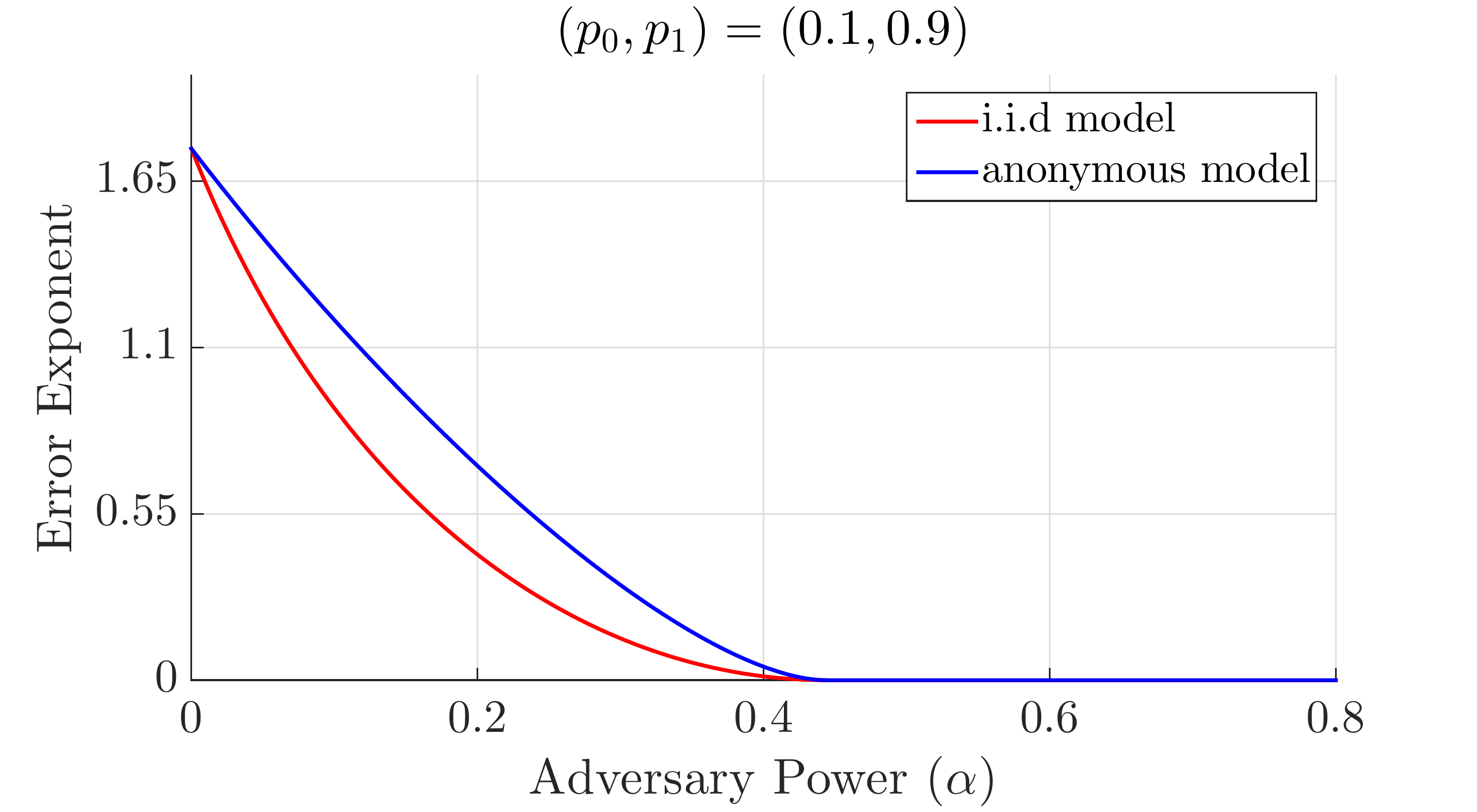}%
	\caption{Comparison between i.i.d. and our setting }%
	\label{fig:Byzantine}
\end{figure}

We see that the achievable type-II error exponent \eqref{eq:byzantine_ach} in our setting is always greater than that in the i.i.d. scenario \eqref{eq:byzantine_iid} (and is \emph{strictly} larger for some $\alpha$) due to the convexity of KL divergence. 
This implies the i.i.d. mixture model \cite{MaranoMatta_09} might be too pessimistic. 
 Figure~\ref{fig:Byzantine} shows a numerical evaluation. 

%% file: sec_proof_optimal.tex
Before proving Theorem~\ref{thm:optimal_test}, let us introduce some definitions that help the exposition. 
\begin{definition}[Ordering Map]\label{def:Pi}
	The ordering map $\Pi(\cdot) : (\mcal{X}^n, \mcal{F}^{\otimes n}) \ra \lp\tilde{\mcal{X}}^n, \tilde{\mcal{F}}\rp$, where $\tilde{\mcal{X}}^n$ is from \eqref{eq:type_space} and $ \tilde{\mcal{F}} \eqDef  \mcal{F}^{\otimes n}\cap \tilde{\mcal{X}}^n$, is  defined as follows:
	$$ \Pi(x^n) \eqDef  (x_{i_1}, x_{i_2},...,x_{i_n}), \text{ such that } x_{i_1}\geq x_{i_2}\geq...\geq x_{i_n}.$$
	The measurability of $\Pi$ is easy to check. 
\end{definition}
\begin{remark}
	If $|\mcal{X}| < \infty$, the mapping $\Pi$ maps a sample $x^n$ to its type, and the space $\tilde{\mcal{X}}^n$ is equivalent to $\mcal{P_X}$.
\end{remark}
\begin{remark}
	We will use $\Pi^{-1}$ to denote the \emph{pre-image} of $\Pi$. That is, for all $\tilde{\mcal{E}} \subseteq \tilde{\mcal{X}}^n$,
	$$ \Pi^{-1}\lp\tilde{\mcal{E}}\rp \eqDef \lbp x^n \in\mcal{X}^n \mid \Pi(x^n)\in\tilde{\mcal{E}} \rbp.$$
	Notice that the measurability of $\Pi$ implies for any $\tilde{\mcal{E}}\in\tilde{\mcal{F}}$, we have $\Pi^{-1}\lp\tilde{\mcal{E}}\rp\in\mcal{F}^{\otimes n}$.
\end{remark}

\begin{definition}[Symmetric Test]\label{def:sym_test}
	We say a test $\phi(x^n)$ is \emph{symmetric}, if it is $\sigma(\Pi(X^n))$-measurable, that is, it can be represented as a composition
	$$ \phi(x^n) = \tilde{\phi}\circ\Pi(x^n), $$
	for some measurable function $\tilde{\phi}: \tilde{\mcal{X}}^n \ra [0,1]$. This implies the test $\phi$ maps a sequence of observations $x^n$ and all its permutations to the same value.
\end{definition}

\begin{lemma}\label{lemma_1}
	Among all symmetric test, $\phi^*(x^n)$, as defined in \eqref{eq:optimal_test}, is optimal.
\end{lemma}
\begin{IEEEproof}[proof of Lemma~\ref{lemma_1}]
	To show the optimality of $\phi^*$, we first transform the original composite hypothesis testing problem to another one in the auxiliary space $\tilde{\mcal{X}}^n$ through the ordering mapping $\Pi(\cdot)$, which turns out to be a simple hypothesis testing problem. Hence, applying Neyman-Pearson lemma, we obtain the optimal test. See Figure~\ref{fig:mapping} for illustration of the relation between the original space and the auxiliary space.
	
	\begin{figure}[htbp]
		\centering
		\includegraphics[width=0.3\linewidth]{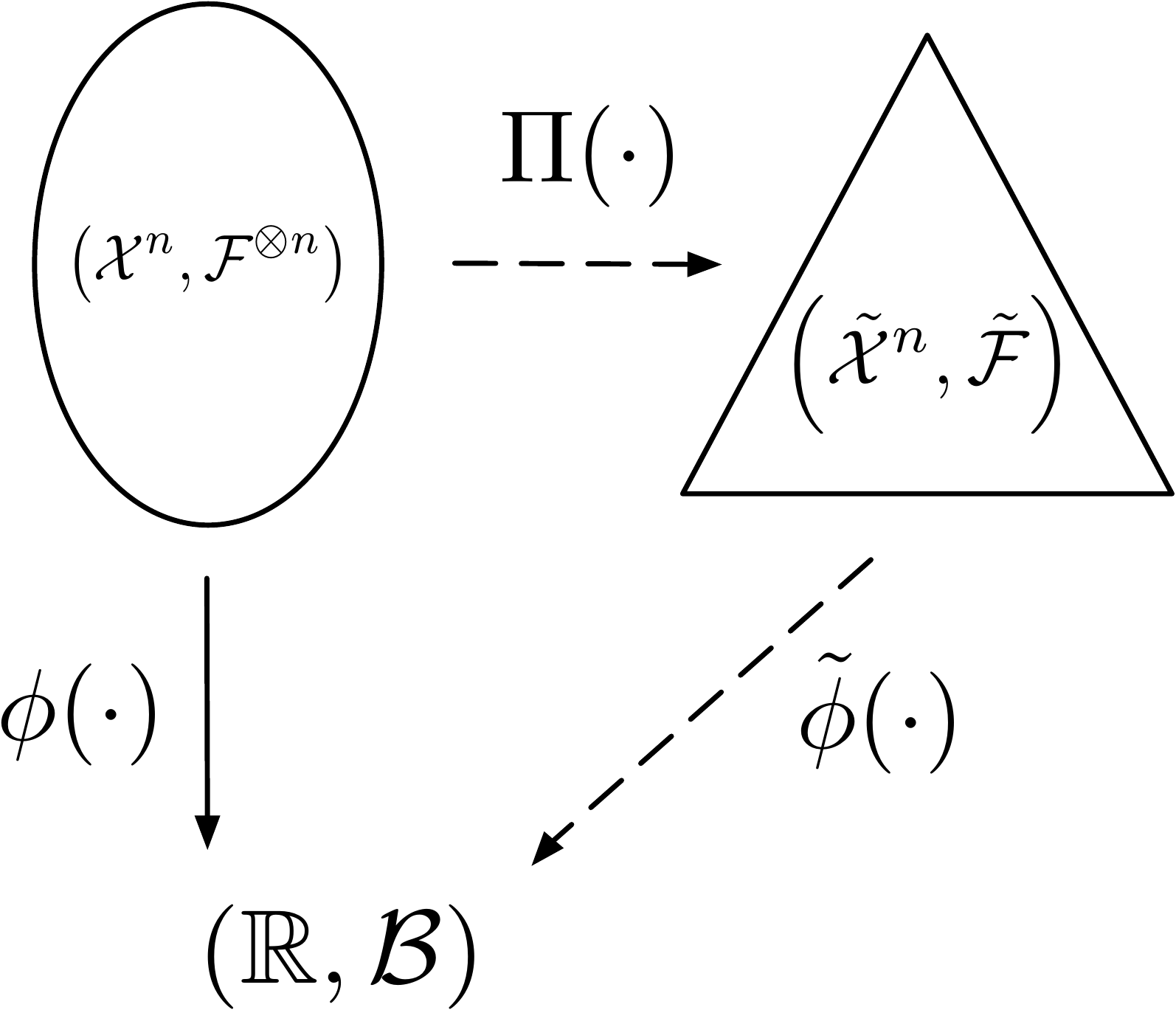} %
		\caption{Illustration of the auxiliary space }%
		\label{fig:mapping}
	\end{figure}
	
\begin{proofpart}
		First, we claim that for all $\sigma \in \mcal{S}_{n,\bm{\nu}}$, the probability measure $ \mbb{P}_{0;\sigma}\circ \Pi^{-1}$, defined on $(\tilde{\mcal{X}}^n, \tilde{\mcal{F}})$, does not depend on $\sigma$ anymore. Thus we can define the probability measure $\tilde{\mbb{P}}_{0} \eqDef \mbb{P}_{0;\sigma}\circ \Pi^{-1}$, such that for all $\sigma$, 
		$$ \lp \mbb{P}_{0;\sigma}, \mcal{F}^{\otimes n}, \mcal{X}^n \rp \overset{\Pi(\cdot)}{\lgra}  \lp \tilde{\mbb{P}}_{0}, \tilde{\mcal{F}}, \tilde{\mcal{X}}^n \rp.$$
		This claim is quite intuitive, since the labeling $\sigma$ corresponds to the order of observations, and the ordering map removes the order.
		
		To show this claim, we first observe that for all $\mcal{E} \in \tilde{\mcal{F}}$,  its pre-image
		\begin{equation}\label{eq:preimage}
			\Pi^{-1}(\mcal{E}) = \bigcup_{\tau \in \mcal{S}_n} \mcal{E}_\tau,
		\end{equation}
		where 
		$\mcal{E}_\tau\eqDef \lbp  (x_{\tau(1)},...,x_{\tau(n)}) \mid (x_1,...,x_n)  \in \mcal{E} \rbp$. 
		Therefore, for any two $\sigma, \sigma' \in \mcal{S}_{n,\bm{\nu}}$, we can write $\sigma' = \pi\circ \sigma$ for some $\pi \in \mcal{S}_n$, and thus have
		\begin{align*}
			\mbb{P}_{0;\sigma}\circ \Pi^{-1}\lbp \mcal{E} \rbp 
			=& \mbb{P}_{0;\sigma}\lbp\bigcup_{\tau \in \mcal{S}_n} \mcal{E}_\tau \rbp
			\overset{\text{(a)}}{=} \mbb{P}_{0;\sigma}\lbp\bigcup_{\tau \in \mcal{S}_n} \mcal{E}_{\tau\circ\pi} \rbp\\
			=& \mbb{P}_{0;\pi\circ\sigma}\lbp\bigcup_{\tau \in \mcal{S}_n} \mcal{E}_\tau \rbp
			= \mbb{P}_{0;\sigma'}\circ \Pi^{-1}\lbp \mcal{E}\rbp,
		\end{align*}
		where the equality (a) holds due to the following fact:
		$$ \forall \pi\in\mcal{S}_n, \, \mcal{S}_n\circ\pi \eqDef \lbp \tau\circ\pi \mid \tau\in\mcal{S}_n\rbp = \mcal{S}_n. $$
		Following the same argument, $\tilde{\mbb{P}}_{1} \eqDef \mbb{P}_{1;\sigma}\circ \Pi^{-1}$ does not depend on $\sigma$ either.
\end{proofpart}
\begin{proofpart}
	Second, let us we consider an auxiliary hypothesis testing problem on $\tilde{\mcal{X}}^n$:
	\begin{equation}\label{eq:aux_test}
		\begin{cases}
		\tilde{\mcal{H}}_0: Z \sim \tilde{\mbb{P}}_0\\
		\tilde{\mcal{H}}_1: Z \sim \tilde{\mbb{P}}_1,
		\end{cases}
	\end{equation}
	and let $\tilde{\phi} : \tilde{\mcal{X}}^n\ra [0,1]$ be a test with type-I and type-II error probabilities as follows:
	\begin{equation*}
		\begin{cases}
		\mathsf{P_F}(\tilde{\phi})\eqDef \E_{\tilde{\mbb{P}}_0}\lb\tilde{\phi}(Z)\rb\\
		\mathsf{P_M}(\tilde{\phi})\eqDef \E_{\tilde{\mbb{P}}_1}\lb 1-\tilde{\phi}(Z)\rb.
		\end{cases}
	\end{equation*}
	We claim that for any symmetric test $\phi(x^n) = \tilde{\phi}\lp\Pi(x^n)\rp$ as defined in Definition~\ref{def:sym_test}, the following holds:
	\begin{equation*}
		\begin{cases}
		\mathsf{P_F}(\tilde{\phi})=\mathsf{P_F}(\phi)\\
		\mathsf{P_M}(\tilde{\phi})=\mathsf{P_M}(\phi).
		\end{cases}
	\end{equation*} 
	To show this, note that a direct calculation gives 
	\begin{align*}
		\mathsf{P_F}(\phi) 
		& = \max_\sigma \E_{\mbb{P}_{0;\sigma}}\lb\phi(X^n)\rb \\
		& = \max_\sigma \E_{\mbb{P}_{0;\sigma}}\lb\tilde{\phi}\lp\Pi(X^n)\rp\rb \\
		& = \max_\sigma \int \tilde{\phi}\lp\Pi(x^n)\rp \mbb{P}_{0;\sigma}(dx^n)\\
		& = \max_\sigma \int \tilde{\phi}\lp z \rp \mbb{P}_{0;\sigma}(\Pi^{-1}(dz))\\
		& = \E_{\tilde{\mbb{P}}_0}\lb\tilde{\phi}(Z)\rb = \mathsf{P_F}(\tilde{\phi}).
	\end{align*}
	For the same reason, $\mathsf{P_M}(\phi) = \mathsf{P_M}(\tilde{\phi})$. 
	Therefore, for any symmetric test on $\mcal{X}^n$,  the corresponding $\tilde{\phi}$ has exactly the same type-I and type-II error probability. Notice that the auxiliary hypothesis testing problem \eqref{eq:aux_test} is simple, so by Neyman-Pearson lemma, we have readily seen that the optimal symmetric test on the original problem should be 
	\begin{equation*}
	\phi^*(x^n) = 
	\begin{cases}
	 1, & \text{ if } \ell'(x^n) > \tau \\
	 \gamma, & \text{ if } \ell'(x^n) = \tau \\
	 0, & \text{ if } \ell'(x^n) < \tau,
	\end{cases}
	\end{equation*}
	where $\ell'(x^n)$ is defined as 
	$$  \ell'(x^n) = \frac{\tilde{\mbb{P}}_1\lp\Pi(x^n)\rp}{\tilde{\mbb{P}}_0\lp\Pi(x^n)\rp}
	 = \frac{\mbb{P}_{1;\sigma}\lbp \Pi^{-1}\lp\Pi(x^n)\rp \rbp}{\mbb{P}_{0;\sigma}\lbp \Pi^{-1}\lp\Pi(x^n)\rp \rb}. $$
\end{proofpart}
\begin{proofpart}
	Finally, we show that $\ell'(x^n)$ is indeed the mixture likelihood ratio $\ell(x^n)$, as defined in \eqref{eq:mlr}. With a slight abuse of notation, let $\Pi_{x^n} \eqDef \Pi^{-1}\lp\Pi(x^n)\rp = \lbp x_{\tau(1)},...,x_{\tau(n)} \mid \tau \in \mcal{S}_n \rbp$. In words, $\Pi_{x^n}$ is the collection of $x^n$ and all its permutations. We observe that 
	\begin{align*}
		\mbb{P}_{1;\sigma}\lbp \Pi^{-1}\lp\Pi(x^n)\rp \rbp 
		& = \sum_{y^n \in \Pi_{x^n}} \mbb{P}_{1;\sigma}\lp y^n \rp\\
		& \overset{\text{(a)}}{=} \lp \sum_{\tau \in \mcal{S}_n} \mbb{P}_{1;\sigma}\lp \tau(x^n) \rp\rp c_1(x^n) \\
		& \overset{\text{(b)}}{=} \lp \sum_{\sigma' \in \mcal{S}_{n,\bm{\nu}}} \mbb{P}_{1;\sigma'}\lp x^n \rp\rp c_1(x^n)c_2(\sigma).
	\end{align*}
	The constant $c_1(x^n)$ in (a) is due to the fact that $x^n = (x_1,...,x_n)$ might not be all distinct, so summing over the set $\lbp \tau(x^n) \mid \tau \in \mcal{S}_n \rbp$ may count an element $y^n \in \Pi_{x^n}$ multiple times.  Note that if $x^n$ are all distinct, then $c_1(x^n) =1$. (b) holds because $ \mbb{P}_{1;\sigma}(\tau(x^n)) = \mbb{P}_{1;\sigma\circ\tau}(x^n) $ and
	$ \mcal{S}_{n,\bm{\nu}} = \mcal{S}_{n,\bm{\nu}}\circ\tau \eqDef \lbp \sigma\circ\tau \mid \sigma\in\mcal{S}_{n,\bm{\nu}} \rbp$. Again, the summation counts $\sigma$ repeatedly, so we normalize by the constant $c_2(\sigma)$. Following the same reason, 
	$$ \mbb{P}_{0;\sigma}\lbp \Pi^{-1}\lp\Pi(x^n)\rp \rbp  =  
	\lp \sum_{\sigma' \in \mcal{S}_{n,\bm{\nu}}} \mbb{P}_{0;\sigma'}\lp x^n \rp\rp c_1(x^n)c_2(\sigma).$$
	Hence,
	\begin{align*}
		\ell'(x^n) 
		& = \frac{\mbb{P}_{1;\sigma}\lbp \Pi^{-1}\lp\Pi(x^n)\rp \rbp}{\mbb{P}_{0;\sigma}\lbp \Pi^{-1}\lp\Pi(x^n)\rp \rbp}\\
		& = \frac{ \lp\sum_{\sigma' \in \mcal{S}_{n,\bm{\nu}}} \mbb{P}_{1;\sigma'}\lp x^n \rp\rp c_1(x^n)c_2(\sigma)}{ \lp\sum_{\sigma' \in \mcal{S}_{n,\bm{\nu}}} \mbb{P}_{0;\sigma'}\lp x^n \rp\rp c_1(x^n)c_2(\sigma)} \\
		& = \frac{\sum_{\sigma}\mbb{P}_{1;\sigma} (x^n)}{\sum_{\sigma}\mbb{P}_{0;\sigma} (x^n)} = \ell(x^n),
	\end{align*} 
	which establishes the claim.
\end{proofpart}
\end{IEEEproof}

\begin{lemma}\label{lemma_2}
	For any general (measurable) test $\psi(x^n) : \mcal{X}^n \ra [0,1]$, there exists a symmetric test $\phi(x^n)$ whose performance is not worse than $\psi$. That is, 
	\begin{equation}\label{eq:sym_optimal}
		\begin{cases}
		\mathsf{P_F}(\phi)\leq\mathsf{P_F}(\psi)\\
		\mathsf{P_M}(\phi)\leq\mathsf{P_M}(\psi).
		\end{cases}
	\end{equation}
 \end{lemma}
 \begin{IEEEproof}[proof of Lemma~\ref{lemma_2}]
 	With a slight abuse of notation, let $\tau(x^n)$ denote the coordinate-permutation function with respect to $\tau \in \mcal{S}_n$, i.e. $\tau(x^n) = (x_{\tau(1)},...,x_{\tau(n)})$.
 	Then we construct $\phi(x^n)$ as follows:
 	$$ \phi(x^n) \eqDef \frac{1}{n!} \sum_{\tau \in \mcal{S}_n} \psi\circ\tau(x^n). $$
 	We claim the following two facts:
 	\begin{enumerate}
 		\item $\phi(x^n)$ is symmetric, and thus can be written as $\tilde{\phi}\circ\Pi(x^n)$ for some $\tilde{\mathcal{F}}$-measurable $\tilde{\phi}$.
 		\item \eqref{eq:sym_optimal} holds for the constructed $\phi$.
 	\end{enumerate}
 	\setcounter{proofpart}{0}
 	\begin{proofpart}
 		
 		To see that $\phi(x^n) = \tilde{\phi}\circ\Pi(x^n)$, we observe that for any $y^n, z^n \in \Pi^{-1}(\tilde{x}^n)$, there exists a permutation $\pi \in \mcal{S}_n$ such that $y^n = \pi(z^n)$. Hence it suffices to verify that for all $\pi \in \mcal{S}_n$, $\phi(x^n) = \phi(\pi(x^n))$.
 		\begin{align*}
 			\phi(\pi(x^n)) 
 			& = \frac{1}{n!} \sum_{\tau \in \mcal{S}_n} \psi\circ\tau\lp \pi (x^n)\rp\\
 			& = \frac{1}{n!} \sum_{\tau \in \mcal{S}_n} \psi\circ\tau\circ \pi(x^n) \\
 			& \overset{\text{(a)}}{=} \frac{1}{n!} \sum_{\tau' \in \mcal{S}_n} \psi\circ\tau'\lp x^n\rp
 			= \phi(x^n).
 		\end{align*}
 		The equality (a) holds due to the fact that
 		$$ \mcal{S}_n\circ\pi \eqDef \lbp \tau\circ\pi \mid \tau \in \mcal{S}_n \rbp = \mcal{S}_n.$$
 		Therefore, $\phi(x^n)$ can be decomposed into  $\tilde{\phi}\circ\Pi(x^n)$. 
 		
 		Next, we check the measurability of $\tilde{\phi}$. Notice that $\phi$ is $\mcal{F}^{\otimes }$-measurable, since both $\psi$ and $\tau$ are measurable. The measurability of $\tau$ follows from the $\tau$-permuted closedness assumption of $\mcal{F}^{\otimes n}$:
 		$$\forall \mcal{A} \in \mcal{F}^{\otimes n}, \mcal{A}_\tau \eqDef \lbp \tau(x^n) \mid x^n \in \mcal{A} \rbp\in \mcal{F}^{\otimes n}.$$
 		Observe that for all Borel-measurable set $\mcal{B}$, we have 
 		\begin{align*}
 			\phi^{-1}\lbp\mcal{B}\rbp = \Pi^{-1}\lbp \tilde{\phi}^{-1} \lbp \mcal{B} \rbp \rbp \in \mcal{F}^{\otimes n} 
 			\Leftrightarrow \bigcup_{\tau\in \mcal{S}_n} \mcal{E}_\tau \in \mcal{F}^{\otimes n},
 		\end{align*}
 		where we use $\mcal{E}$ to denote event $\tilde{\phi}^{-1} \lbp \mcal{B} \rbp$, and $\mcal{E}_\tau$ to denote the $\tau$-permuted event of $\mcal{E}$, as defined in \eqref{eq:tau_shift}. Notice here we use the fact given by \eqref{eq:preimage}. Therefore it suffices to check 
 		$$ \forall \mcal{E} \subseteq \tilde{\mcal{X}}^n, \bigcup_{\tau\in \mcal{S}_n} \mcal{E}_\tau \in \mcal{F}^{\otimes n} \Rightarrow \mcal{E} \in \mcal{F}^{\otimes n}\cap \tilde{\mcal{X}}^n=\tilde{\mcal{F}}.$$
 		We claim that indeed, 
 		$$ \lbp \bigcup_{\tau\in \mcal{S}_n} \mcal{E}_\tau\rbp \cap \tilde{\mcal{X}}^n = \mcal{E}, $$
 		for every $\mcal{E} \subseteq \tilde{\mcal{X}}^n$.
 		This is because 
 		\begin{enumerate}
 			\item Since $ \mcal{E} \subseteq  \tilde{\mcal{X}}^n,$ we have 
 			$ \mcal{E} = \mcal{E}\cap  \tilde{\mcal{X}}^n \subseteq \lbp\bigcup_{\tau\in \mcal{S}_n} \mcal{E}_\tau\rbp \cap \tilde{\mcal{X}}^n$. 
 			\item For any $\tau$ and for any $x^n \in \mcal{E}_\tau \cap \tilde{\mcal{X}}^n$, $x^n \in \mcal{E}$. Hence,
 			$\forall \tau \in \mcal{S}_n, \,\mcal{E}_\tau \cap \tilde{\mcal{X}}^n \subseteq \mcal{E}$, that is, 
 			$ \lbp \bigcup_{\tau\in \mcal{S}_n} \mcal{E}_\tau\rbp \cap \tilde{\mcal{X}}^n \subseteq \mcal{E}$. 
 		\end{enumerate}
 		Hence,  
 		$$ 
		\bigcup_{\tau\in \mcal{S}_n} \mcal{E}_\tau \in \mcal{F}^{\otimes n}
		\implies \mcal{E} =  \lbp \bigcup_{\tau\in \mcal{S}_n} \mcal{E}_\tau\rbp \cap \tilde{\mcal{X}}^n \in \mcal{F}^{\otimes n} \cap \tilde{\mcal{X}}^n = \tilde{\mcal{F}},$$
 		showing that $\tilde{\phi}$ is $\tilde{\mcal{F}}-$measurable.
 	\end{proofpart}
 	\begin{proofpart}
 		We show that $\phi(x^n)$ cannot be worse than $\psi(x^n)$. Observe that for all $\tau \in \mcal{S}_n$, we have 
 		\begin{align*}
 			\mathsf{P_F}(\psi\circ\tau) 
 			& =\max_{\sigma\in\mcal{S}_{n,\bm{\nu}}} \E_{\mbb{P}_{0;\sigma}}\lb \psi\lp\tau(X^n) \rp\rb
 			 = \max_{\sigma\in\mcal{S}_{n,\bm{\nu}}} \E_{\mbb{P}_{0;\sigma\circ\tau^{-1}}}\lb \psi(X^n) \rb
 			 = \max_{\sigma'\in\mcal{S}_{n,\bm{\nu}}} \E_{\mbb{P}_{0;\sigma'}}\lb \psi(X^n) \rb
 			= \mathsf{P_F}(\psi).
 		\end{align*}
 		Again, the third equality holds due to the fact
 		$$ \mcal{S}_{n,\bm{\nu}}\circ\tau^{-1} \eqDef \lbp \sigma\circ\tau^{-1} \mid \sigma \in \mcal{S}_{n,\bm{\nu}} \rbp = \mcal{S}_{n,\bm{\nu}}. $$
 		Therefore, we have
 		\begin{align*}
 			\mathsf{P_F}(\phi) = 
 			& \max_{\sigma} \E_{\mbb{P}_{0;\sigma}}\lb \frac{1}{n!}\sum_{\tau\in \mcal{S}_n}\psi\circ\tau(X^n) \rb\\
 			\leq & \frac{1}{n!}\sum_{\tau\in \mcal{S}_n}\max_{\sigma} \E_{\mbb{P}_{0;\sigma}}\lb \psi\circ\tau(X^n) \rb\\
 			= & \frac{1}{n!}\sum_{\tau\in \mcal{S}_n} \mathsf{P_F}(\psi\circ\tau)= \mathsf{P_F}(\psi).
 		\end{align*}
 		Following the same argument, we obtain $\mathsf{P_M}(\phi) \leq \mathsf{P_M}(\psi)$, and the proof completes.
 	\end{proofpart}
 \end{IEEEproof}
 Finally, the proof of Theorem~\ref{thm:optimal_test} directly follows from Lemma~\ref{lemma_1} and Lemma~\ref{lemma_2}.
 \begin{IEEEproof}[Proof of Theorem~\ref{thm:optimal_test}]
 	From Lemma~\ref{lemma_2}, we only need to consider symmetric tests. From Lemma~\ref{lemma_1}, we see that the optimal test among all symmetric tests is the mixture likelihood test, as defined in \eqref{eq:optimal_test}. This establishes Theorem~\ref{thm:optimal_test}.
 \end{IEEEproof}
 \begin{remark}
 Notice that in the above proof, we do not make use of assumptions on the distribution of $X^n$, such as independence. Indeed, the proof indicates that for the anonymous composite hypothesis testing problem, under the minimax criterion (i.e. to minimize the worst case error), we should always design tests based on the empirical distribution of $X^n$ (i.e. as a function of $\Pi(x^n)$). This principle also holds for other statistical inference problems, such as $M$-ary hypothesis testing.   
 \end{remark}

%% file: sec_proof_asymptotics.tex
 	For the case $|\mcal{X}|<\infty$, the auxiliary space $\tilde{\mcal{X}}$ is equivalent to the space of all probability measures on $\mcal{X}$, that is, $\mcal{P_X}$, and the mapping $\Pi(x^n)$ maps a sequence of samples to its type $\Pi_{x^n}$. According to Lemma~\ref{lemma_2}, the optimal test is symmetric, which implies that we only need to consider tests depending on the type. For tests depending only on the empirical distribution, it is natural to view their acceptance region as a collection of empirical distribution, that is, a (measurable) subset of $\mcal{P_X}$. This motivates us to apply Sanov's theorem. We begin with the following generalization of Sanov's result:

	\begin{lemma}[Generalized Sanov Theorem]\label{lemma:general_sanov}
		Let $\lba \mcal{X} \rba < \infty$, and $\Gamma \subseteq \mcal{P_X}$ be a collection of distributions on $\mcal{X}$. Then for all $\sigma \in \mcal{S}_{n,\bm{\nu}}$ and $\theta \in \{0,1\}$, we have 
		\begin{align}\label{eq:general_sanov}
 			&-\inf_{\substack{ {[U_1\ ...\ U_K]^\intercal\in\lp\mcal{P_X}\rp^K} \\ \bm{\alpha}^\intercal\bm{U}\in\text{int }\Gamma}} \sum_{k=1}^K \alpha_k\KLD{U_k}{P_{\theta;k}} \\
 			\leq &\liminf_{n\ra\infty} \frac{1}{n}\log \mbb{P}_{\theta;\sigma}\lbp \Pi_{x^n} \in \Gamma \rbp \\
 			\leq & \limsup_{n\ra\infty} \frac{1}{n}\log \mbb{P}_{\theta;\sigma}\lbp \Pi_{x^n} \in \Gamma \rbp \\
 			\leq & -\inf_{\substack{ {[U_1\ ...\ U_K]^\intercal\in\lp\mcal{P_X}\rp^K} \\ \bm{\alpha}^\intercal\bm{U}\in\text{cl }\Gamma}} \sum_{k=1}^K \alpha_k\KLD{U_k}{P_{\theta;k}},
 		\end{align}
		where in taking the limits, we assume that $\lim_{n\ra\infty}\frac{n_k}{n} = \alpha_k$, for all $k=1,...,K$.
		In particular, if the infimum in the right-hand side is equal to the infimum in the left-hand side, then we have 
		$$ \lim_{n\ra\infty} \frac{1}{n}\log \mbb{P}_{\theta;\sigma}\lbp \Pi_{x^n} \in \Gamma \rbp 
		= -\inf_{\substack{ {[U_1\ ...\ U_K]^\intercal\in\lp\mcal{P_X}\rp^K} \\ \bm{\alpha}^\intercal\bm{U}\in\text{cl }\Gamma}} \sum_k	 \alpha_k\KLD{U_k}{P_{\theta;k}}. $$
	\end{lemma}
	The proof is a direct extension of Lemma~\ref{lemma:sanov}, except that we replace the i.i.d. measure with the product of independent non-identical ones, $\mbb{P}_{\theta;\sigma}$. For the detailed proof, please refer to Appendix~\ref{sec:proof_general_sanov}.

Motivated by the generalized Sanov Theorem, we further define the following generalized divergence to measure \emph{how far} from one set of distributions $\bm{Q} \eqDef [Q_1\ ...\ Q_K]^\intercal$ to another set of distributions $\bm{P} \eqDef [P_1\ ...\ P_K]^\intercal$:

\begin{definition}\label{def:d}
{Let $\bm{P} = [P_1\ ...\ P_K]^\intercal$ and $\bm{Q} = [Q_1\ ...\ Q_K]^\intercal$ are both in $\lp\mcal{P_X}\rp^K$. Let $\bm{\alpha} = [\alpha_1\ ...\ \alpha_K]^\intercal$ be a $K$-tuple probability vector. Define 
\begin{equation}\label{eq:distance}
			\begin{split}
					D_{\bm{\alpha}}(\bm{P};\bm{Q}) \eqDef &\inf\limits_{\bm{U} \in (\mcal{P_X})^K}\, \sum\nolimits_{k=1}^K\alpha_k\KLD{U_k}{Q_k}\\
			& \begin{array}{ll}
			\text{subject to} &\bm{\alpha}^\intercal\bm{U}=\bm{\alpha}^\intercal\bm{P}
			\end{array}
			\end{split}.
\end{equation}
}		
\end{definition}

	Thus \eqref{eq:general_sanov} in Lemma~\ref{lemma:general_sanov} can be rewritten as 

	$$ -\inf_{\bm{\alpha}^\intercal\bm{U} \in\text{int }\Gamma} D_{\bm{\alpha}}(\bm{U}; \bm{P}_\theta) 
 			\leq \liminf_{n\ra\infty} \frac{1}{n}\log \mbb{P}_{\theta;\sigma}\lbp \Pi_{x^n} \in \Gamma \rbp 
 			\leq  \limsup_{n\ra\infty} \frac{1}{n}\log \mbb{P}_{\theta;\sigma}\lbp \Pi_{x^n} \in \Gamma \rbp 
 		\leq  -\inf_{\bm{\alpha}^\intercal\bm{U} \in\text{cl }\Gamma} D_{\bm{\alpha}}(\bm{U}; \bm{P}_\theta). $$	
 	Also, the result of Theorem~\ref{thm:asymptotics}, \eqref{eq:lowerbd}, is equivalent to the following statement:
	$$ E^*(\epsilon,\bm{\alpha}) = D_{\bm{\alpha}}(\bm{P}_0;\bm{P}_1).$$
\begin{remark}
	Intuitively, $D_{\bm{\alpha}}(\bm{P};\bm{Q})$ measures how far between $\bm{P}$ and $\bm{Q}$. However, $D_{\bm{\alpha}}(\cdot;\cdot)$ is not a divergence, since $D_{\bm{\alpha}}(\bm{P};\bm{Q}) = 0$ does not always imply $\bm{P} = \bm{Q}$.
\end{remark}
Notice that for any fixed $\bm{Q}\in\lp\mcal{P_X}\rp^K$, $D_{\bm{\alpha}}(\bm{P};\bm{Q})$ can be regarded as a function of $\bm{P}$. Moreover, this function depends only on the mixture of $\bm{P}$, say, $\bm{\alpha}^\intercal\bm{P}$. Therefore, for notional convenience, let us use $f_{\bm{Q}}(\cdot):\mcal{P_X}\ra \mbb{R}\cup\{+\infty\}$ to denote this function:
\begin{equation*}
			\begin{split}
					f_{\bm{Q}}(T) \eqDef &\inf\limits_{\bm{U} \in (\mcal{P_X})^K}\, \sum\nolimits_{k=1}^K\alpha_k\KLD{U_k}{Q_k}\\
			& \begin{array}{ll}
			\text{subject to} &\bm{\alpha}^\intercal\bm{U}=T
			\end{array}
			\end{split}.
\end{equation*}
In other words, 
$$ f_{\bm{Q}}(\bm{\alpha}^\intercal\bm{P}) = D_{\bm{\alpha}}(\bm{P};\bm{Q}).$$

Before entering the main proof of Theorem~\ref{thm:asymptotics}, let us introduce some properties of $f_{\bm{Q}}(\cdot)$.

\begin{lemma}\label{lemma:property_d}
	Let $\bm{Q} \in \lp\mcal{P_X}\rp^K$ and $f_{\bm{Q}}(\cdot): \mcal{P_X} \ra \mbb{R}\cup\lbp+\infty\rbp$ be defined as Definition~\ref{def:d} and above. Then, 
	\begin{enumerate}
		\item $f_{\bm{Q}}(\bm{\alpha}^\intercal\bm{Q}) = 0$
		\item The collection of all $T\in\mcal{P_X}$ such that $f_{\bm{Q}}(T) < \infty$, denoted as 
			$$ \mcal{C}_{\bm{Q}} \eqDef \lbp T\in\mcal{P_X} : f_{\bm{Q}}(T) < \infty \rbp, $$
		 	is a \emph{compact, convex} subset of $\mcal{P_X}$.
		\item $f_{\bm{Q}}(T)$ is a \emph{convex, continuous} function of $T$ on $\mcal{C}_{\bm{Q}}$ (and by the compactness of $\mcal{C}_{\bm{Q}}$, $f_{\bm{Q}}(T)$ is also \emph{uniformly} continuous).
	\end{enumerate}
\end{lemma}
Proof of Lemma~\ref{lemma:property_d} can be found in Appendix~\ref{sec:proof_property_d}.

\begin{IEEEproof}[proof of Theorem~\ref{thm:asymptotics}]
 	\setcounter{proofpart}{0}
 	\begin{proofpart}[Achievability]
 		Let $\delta > 0$ and consider the test : 
 		$$\phi(x^n) \eqDef \mathbbm{1}_{\lbp x^n: \KLD{\Pi_{x^n}}{M_0(\bm{\alpha})} > \delta \rbp}. $$
 		Denote the acceptance region of $\phi$ as $\Gamma \eqDef \lbp T \in \mcal{P_X}: \KLD{T}{M_0(\bm{\alpha})} > \delta \rbp$.
 		Then the exponent of type-I error probability $\mathsf{P_F}(\phi)$ can be bounded by
 		\begin{align*}
 		&\liminf_{n\ra\infty} \frac{1}{n}\log \E_{\Pnull} \lb \phi(X^n) \rb \\
 		& =\liminf_{n\ra\infty} \frac{1}{n}\log\Pnull \lbp \Pi_{x^n} \in \Gamma \rbp \\
 		&\overset{\text{(a)}}{\geq} \inf_{ T \in\text{cl }\Gamma} f_{\bm{P}_0}(T)\\ 
 		&\overset{\text{(b)}}{\geq}\delta,
 		\end{align*}
 		where (a) holds by Lemma~\ref{lemma:general_sanov}, and be holds due to the the convexity of KL divergence: 
 		\begin{equation*}
			\begin{split}
					\KLD{T}{M_0(\bm{\alpha})} \leq &\min\limits_{\bm{U} \in (\mcal{P_X})^K}\, \sum\nolimits_{k=1}^K\alpha_k\KLD{U_k}{P_{0;k}} = f_{\bm{P}_0}(T) \\
			& \begin{array}{ll}
			\text{subject to} &\bm{\alpha}^\intercal\bm{U}=T 
			\end{array}
			\end{split}.
		\end{equation*}
 		Notice that for any $\delta > 0$, as $n$ large enough, we must have 
 		$$ \mathsf{P_F}(\phi) < \epsilon. $$
 		On the other hand, the exponent of type-II error probability $\ul{E}^*(\epsilon,\bm{\alpha})$ can be bounded by 
 		\begin{align}
 		&\liminf_{n\ra\infty} \frac{1}{n}\log \E_{\Palt} \lb \phi(X^n) \rb \nonumber \\
 		& =\liminf_{n\ra\infty} \frac{1}{n}\log\Palt \lbp X^n: \KLD{\Pi_{X^n}}{M_0(\bm{\alpha})} \leq \delta \rbp \nonumber \\
 		&\geq \inf_{ T \in\text{cl }\Gamma^c} f_{\bm{P}_1}(T), \label{eq:gamma_c}
 		\end{align}
 		By Pinsker's inequality (Theorem~6.5 in \cite{PolWu17}), we have
 		$$ \text{cl}\lp\Gamma^c\rp = \lbp T \in \mcal{P_X}: \KLD{T}{M_0(\bm{\alpha})} \leq \delta \rbp \subseteq  \lbp T \in \mcal{P_X}: \lVert T-M_0(\bm{\alpha})\rVert_1\leq \sqrt{2\delta} \rbp \eqDef B_{\sqrt{2\delta}}(M_0(\bm{\alpha})),$$
 		so \eqref{eq:gamma_c} can be further lower bounded by 
 		$$ \inf_{ T \in\text{cl }\Gamma^c} f_{\bm{P}_1}(T) \geq \inf_{ T \in B_{\sqrt{2\delta}}(M_0(\bm{\alpha}))} f_{\bm{P}_1}(T). $$
 		Also, by the continuity (Lemma~\ref{lemma:property_d}) of $f_{\bm{P}_1}(\cdot)$, 

 		$$ \inf_{ T \in B_{\sqrt{2\delta}}(M_0(\bm{\alpha}))} f_{\bm{P}_1}(T) = f_{\bm{P}_1}\lp M_0(\bm{\alpha})\rp+\Delta(\delta),$$
 		with 
 		$$ \lim_{\delta\ra 0} \Delta(\delta) = 0. $$
 		Finally, since $\delta$ can be chosen arbitrarily small, we have 
		\begin{equation}\label{eq:lowerbd_n}
 		\ul{E}^*(\epsilon,\bm{\alpha}) \ge f_{\bm{P}_1}\lp M_0(\bm{\alpha})\rp = D_{\bm{\alpha}}\lp \bm{P}_0;\bm{P}_1\rp.
 		\end{equation}
 	\end{proofpart} 
 	\begin{proofpart}[Converse]
We have shown that symmetric test is optimal in Lemma~\ref{lemma_2}. Hence, in the following, it suffices to consider symmetric tests. 

 		For an arbitrary symmetric test $\psi : \mcal{P}_n\ra [0,1]$ such that its type-I error probability $\mathsf{P_F}(\psi) < \epsilon$, we shall lower bound its type-II error probability as follows. Let $\mcal{A}^{(n)} \eqDef \lbp T\in\mcal{P}_n : \psi(T)\leq 1/2 \rbp$, and recall that 
 		$$ \tilde{\mbb{P}}_0 \eqDef \mbb{P}_{0;\sigma}\circ \Pi^{-1} $$ is a probability measure independent of $\sigma$. Then, we have 
 		\begin{align*}
 			 \epsilon 
 			 &> \E_{\tilde{\mbb{P}}_{0}} \lb \psi(T) \rb 
 			 = \sum_{T\in\mcal{P}_n} \tilde{\mbb{P}}_{0}(T)\psi(T)
 			 \geq  \sum_{T\in\lp\mcal{A}^{(n)}\rp^c} \tilde{\mbb{P}}_{0}(T)\psi(T)\\
 			 &\overset{\text{(a)}}{>} \frac{1}{2}\sum_{T\in\lp\mcal{A}^{(n)}\rp^c} \tilde{\mbb{P}}_{0}(T) 
			 = \frac{1}{2}\lp 1- \tilde{\mbb{P}}_{0} \lbp\mcal{A}^{(n)}\rbp\rp,
 		\end{align*}
 		(a) holds since for all $T\notin \mcal{A}^{(n)}$, $\psi(T) > 1/2$.
 		In other words, we have 
 		$$ \tilde{\mbb{P}}_{0} \lbp\mcal{A}^{(n)}\rbp > 1-2\epsilon. $$
 		On the other hand, let $\mcal{B}^{(n)} \eqDef \lbp T \in \mcal{P}_n \mid \KLD{T}{M_0(\bm{\alpha})} \leq \delta \rbp$. Then, according to the analysis in type-I error probability in the achievability part, we have 
 		$$ \tilde{\mbb{P}}_{0} \lbp\mcal{B}^{(n)}\rbp > 1-\epsilon. $$
 		Applying union bound, we see that 
 		$$ \tilde{\mbb{P}}_{0} \lbp\mcal{A}^{(n)} \cap \mcal{B}^{(n)}\rbp > 1-3\epsilon ,$$ 
 		and hence for $\epsilon < \frac{1}{3}$, $\mcal{A}^{(n)} \cap \mcal{B}^{(n)}$ is non-empty.
 		
 		Let $V_n^* \in \mcal{A}^{(n)} \cap \mcal{B}^{(n)}$ and define $\tilde{\mbb{P}}_{1} \eqDef \mbb{P}_{1;\sigma}\circ \Pi^{-1}$ (which is also independent of $\sigma$). Again we have 
 		\begin{align*}
 			\mathsf{P_F}(\psi) 
 			& = \E_{\tilde{\mbb{P}}_{1}} \lb 1-\psi(T) \rb \\
 			& \geq \sum_{T \in \mcal{A}^{(n)}}\lp 1-\psi(T)\rp \tilde{\mbb{P}}_{1}\lbp T \rbp \\
 			& \geq \frac{1}{2}\tilde{\mbb{P}}_{1}\lbp V_n^* \rbp.
 		\end{align*}
 		We further estimate $\tilde{\mbb{P}}_{1}\lbp V_n^* \rbp$ by 
 		\begin{align*}
 			\tilde{\mbb{P}}_{1}\lbp V_n^* \rbp 
 			=& \Palt\lbp T_n(V_n^*) \rbp \\
 			= &\sum\limits_{\substack{U_k\in\mcal{P}_{n_k}:\\ \sum_k \alpha_k U_k = V_n^*}}\prod_{k=1}^{K} P_{1;k}^{\otimes n_k}\lbp T_{n_k}(U_k) \rbp \\
 			= & \sum\limits_{\substack{U_k\in\mcal{P}_{n_k}:\\ \sum_k \alpha_k U_k = V_n^*}} 2^{-\sum_k n_k \KLD{U_k}{P_{1;k}}}\\
 			\geq & \max\limits_{\substack{U_k\in\mcal{P}_{n_k}:\\ \sum_k \alpha_k U_k = V_n^*}} 2^{-\sum_k n_k \KLD{U_k}{P_{1;k}}}\\
 			= &  2^{-n \tilde{D}_n},
 		\end{align*}
 		where 
 		$$ \tilde{D}_n \eqDef \min\limits_{\substack{U_k\in\mcal{P}_{n_k}:\\ \sum_k \alpha_k U_k = V_n^*}}\lp \sum_k\frac{n_k}{n} \KLD{U_k}{P_{1;k}} \rp.$$
 		Notice that since $V_n^* \in \mcal{B}^{(n)}$, so we have $$ \KLD{V_n^*}{M_0(\bm{\alpha})} \leq \delta. $$
 		Since $\delta$ can be chosen arbitrarily small, as $\delta \ra 0$ and $n \ra \infty$ (with $\frac{n_k}{n}\ra\alpha_k$), we have 
 		\begin{align*} 
		\ol{E}^*(\epsilon,\bm{\alpha}) 
		& \le \lim\limits_{n\ra\infty}\tilde{D}_n \\
		& = \min\limits_{\substack{U_k\in\mcal{P_X}:\\ \sum_k \alpha_k U_k = M_0(\bm{\alpha})}}\lp \sum_k\alpha_k \KLD{U_k}{P_{1;k}} \rp \\
		& = f_{\bm{P}_1}\lp M_0(\bm{\alpha})\rp \\
		& = D_{\bm{\alpha}}\lp \bm{P}_0;\bm{P}_1\rp,
		\end{align*} which completes the proof.
 	\end{proofpart}
 \end{IEEEproof}

%% file: sec_chernoff.tex
So far,  for asymptotic regime, we have been focusing on Neyman-Pearson's formulation, in which we minimize the worst-case type-II error probability, subject to the worst-case type-I error probability not being larger than a constant $\epsilon$. It is natural to extend the result from Section~\ref{sec:results} to Chernoff's regime, where we aim to minimize the average probability of error:
$$ \mathsf{P_e}^{(n)}(\phi) \eqDef \pi_0\mathsf{P_F}^{(n)}+\pi_1\mathsf{P_M}^{(n)}.$$
Note that $\pi_0$ and $\pi_1$ are the prior distributions of $\mcal{H}_0$ and $\mcal{H}_1$ and do not scale with $n$. As suggested by Theorem~\ref{thm:optimal_test}, the optimal test is the mixture likelihood ratio test, so we only need to specify the corresponding threshold $\tau$. However, the mixture likelihood ratio involves summation over $\mcal{S}_{n,\bm{\nu}}$, making the computation complexity extremely high. Even for the case $|\mcal{X}| < \infty$, the computation still takes $\Theta\lp n^{\lba\mcal{X}\rba}\rp$ operations and thus is difficult to implement. To break the computational barrier, we propose an asymptotically optimal test, based on information projection, which achieves the optimal exponent of the average probability of error. Moreover, the result can be generalized to determine the achievable exponent region $\mcal{R}$, the collection of all achievable pairs of exponents:
 $$\mcal{R}\eqDef \lbp (E_0, E_1) \mid \text{ there exists a test }\phi, \text{ such that } 
 \mathsf{P_F}^{(n)}(\phi) \preceq 2^{-n E_0}, \,
\mathsf{P_M}^{(n)}(\phi) \preceq 2^{-n E_1} \rbp,$$
where a sequence $a_n \preceq 2^{-n E_0}$ means $a_n$ decays to zero at the rate faster than $E_0$ , that is,
$$ -\liminf\limits_{n\ra\infty} \frac{1}{n}\log a_n \geq E_0.$$

\subsection{Asymptotically Optimal Test in Chernoff's Regime}

\begin{theorem}[Efficient Test]\label{thm:eff_test}
Recall the function $f_{\bm{P}}(T): \mcal{P_X} \ra \mbb{R}\cup\{+\infty\}$ defined in Defintion~\ref{def:d}.
Consider the following test based on the function $f_{\bm{P}_0}(\cdot)$ and $f_{\bm{P}_1}(\cdot)$:
	\begin{equation}\label{eq:eff_test}
		\phi_\text{eff}(x^n) \eqDef
		\begin{cases}
		 0 , \, \text{ if } f_{\bm{P}_1}(\Pi_{x^n}) > f_{\bm{P}_0}(\Pi_{x^n})\\
		 1 , \, \text{ else } f_{\bm{P}_1}(\Pi_{x^n}) \leq f_{\bm{P}_0}(\Pi_{x^n})
		.\end{cases}
	\end{equation}
	Then $\phi_\text{eff}$ is asymptotically optimal in Chernoff's regime. That is, for all priors $\pi_0, \pi_1$, for all tests $\phi$, and for all $n$ large enough,
	$$ -\frac{1}{n}\log\lp P_e(\phi) \rp \leq  -\frac{1}{n}\log\lp P_e(\phi_\text{eff})\rp.$$
\end{theorem}
\begin{remark}
	From the convexity of KL-divergence and the space $\mcal{P_X}$, the function $f_{\bm{P}}(\cdot)$ is indeed the minimization of a convex function. Hence the proposed test in Theorem~\ref{thm:eff_test} can be computed efficiently.
\end{remark}
\begin{IEEEproof}
Let us set some notations. For each $\bm{P} \in (\mcal{P_X})^K$, we use $B_r(\bm{P}) \subseteq \mcal{P_X}$ to denote the $r$-ball centered at $T$ with respect to $f_{\bm{P}}(\cdot)$:
$$ B_r(\bm{P}) \eqDef \lbp T\in\mcal{P_X} \mid f_{\bm{P}}(T) < r \rbp. $$
By the continuity of $f_{\bm{P}}(\cdot)$ (from Lemma~\ref{lemma:property_d}), $B_r(\bm{P})$ is an open set.
Then, define the largest packing radius between $\bm{P}_0, \bm{P}_1$ as follows: 
$$ r^* \eqDef \sup_r\lbp B_r(\bm{P}_0) \cap B_r(\bm{P}_1) = \emptyset \rbp. $$
\begin{figure}[htbp]
		\centering
		\includegraphics[width=0.35\linewidth]{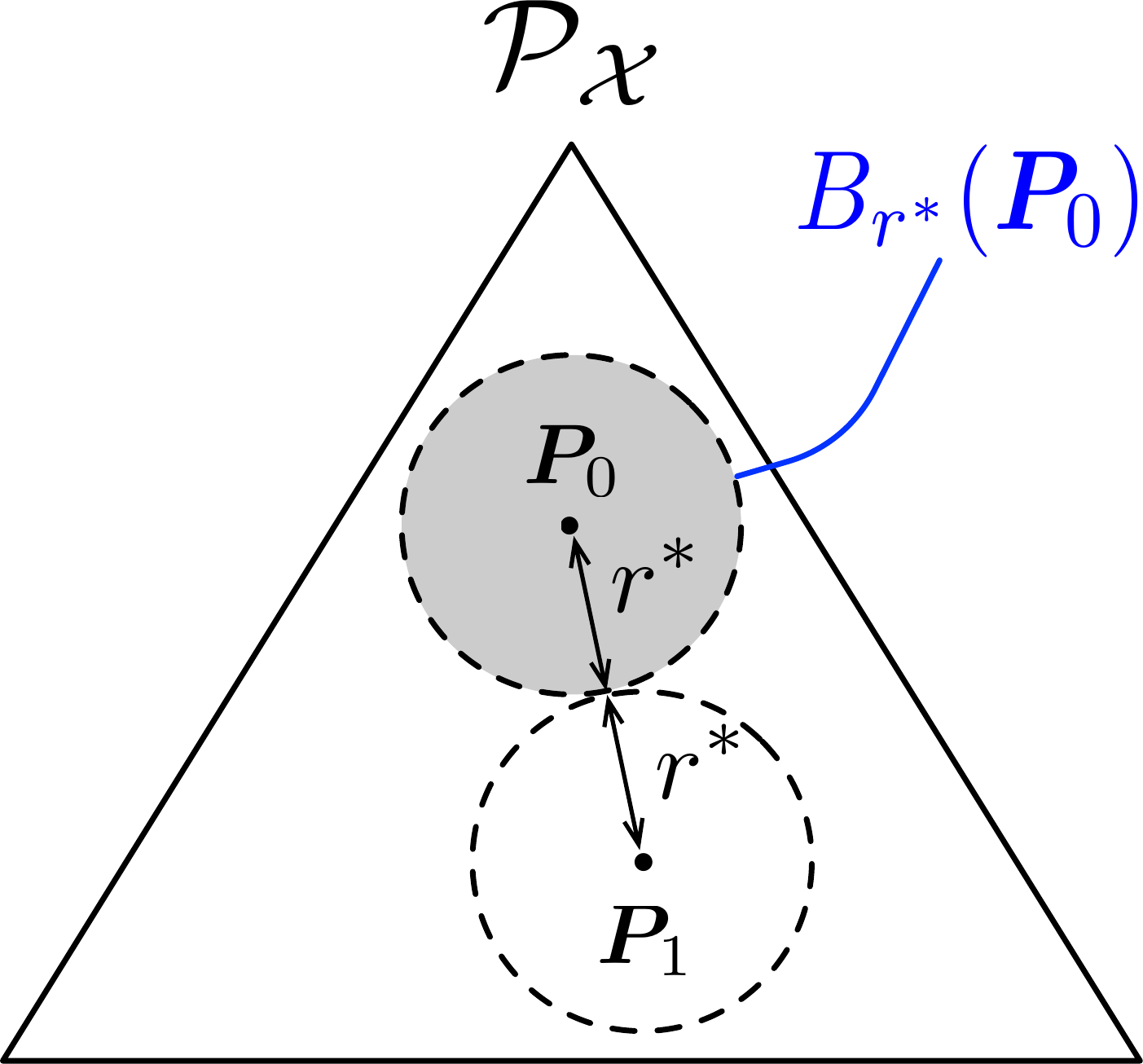} %
		\caption{Illustration of $B_r(\cdot)$ and $r^*$ }%
		\label{fig:B_r}
	\end{figure}
See Figure~\ref{fig:B_r} for illustration.

The rest of the proof will be organized as follows:
we first show that $\phi_\text{eff}$ has error exponent at least $r^*$ (the achievability part):
$$ -\lim_{n\ra \infty}\frac{1}{n}\log\lp P_e(\phi_\text{eff})\rp \geq r^*.$$ Then, we will prove that for all tests, the error exponent will be at most $r^*$ (the converse part).

\setcounter{proofpart}{0}
\begin{proofpart}[Achievability]
Define 
$$ \mcal{A} \eqDef \lbp T\in \mcal{P_X} \mid f_{\bm{P}_1}(T) \leq f_{\bm{P}_0}(T) \rbp,$$
and notice that 
\begin{equation*}
	\begin{cases}
		 \mathsf{P_F}^{(n)}(\phi_\text{eff}) =\mbb{P}_{0;\sigma}\lbp \Pi_{x^n}\in \mcal{A} \rbp \\
		 \mathsf{P_M}^{(n)}(\phi_\text{eff}) =\mbb{P}_{1;\sigma}\lbp \Pi_{x^n}\in \mcal{A}^c \rbp, 
	\end{cases}
\end{equation*}
for any arbitrary $\sigma$ (recall that $\phi_{\text{eff}}$ depends only on the empirical distribution and therefore is symmetrical, so the error is independent of the choice of a specific $\sigma$).

By the generalized Sanov's theorem (Lemma~\ref{lemma:general_sanov}), we see that the exponent of  $\mathsf{P_F}^{(n)}(\phi_\text{eff})$ is lower bounded by 
$ \inf_{T\in\text{cl }\mcal{A}} f_{\bm{P}_0}(T)$. Similarly, the exponent of $\mathsf{P_M}^{(n)}(\phi_\text{eff})$ is lower bounded by 
$ \inf_{T\in\text{cl }\mcal{A}^c} f_{\bm{P}_1}(T)$.
It is not hard to see that indeed,
\begin{equation}\label{eq:chrnff_ach_1}
 \inf_{T\in\text{cl }\mcal{A}} f_{\bm{P}_0}(T) = \inf_{T\in\mcal{A}} f_{\bm{P}_0}(T),
\end{equation}
and 
\begin{equation}\label{eq:chrnff_ach_2}
\inf_{T\in\text{cl }\mcal{A}^c} f_{\bm{P}_1}(T)= \inf_{T\in\mcal{A}^c} f_{\bm{P}_1}(T).
\end{equation}
Equation \eqref{eq:chrnff_ach_1} holds since $\mcal{A}$ is a closed set (it is a pre-image of a continuous function from a closed set), so $\text{cl }\mcal{A} = \mcal{A}$. For the equation \eqref{eq:chrnff_ach_2}, we notice that $\mcal{A}^c$ is open, and hence the infimum of a continuous function on $\mcal{A}^c$ is actually equal to the infimum on $\text{cl }\mcal{A}^c$.

Hence, it suffices to show that 
$$ \inf_{T\in\mcal{A}} f_{\bm{P}_0}(T) \geq r^*, \, \inf_{T\in\mcal{A}^c} f_{\bm{P}_1}(T) \geq r^*.$$ 
\begin{figure}[htbp]
		\centering
		\includegraphics[width=0.35\linewidth]{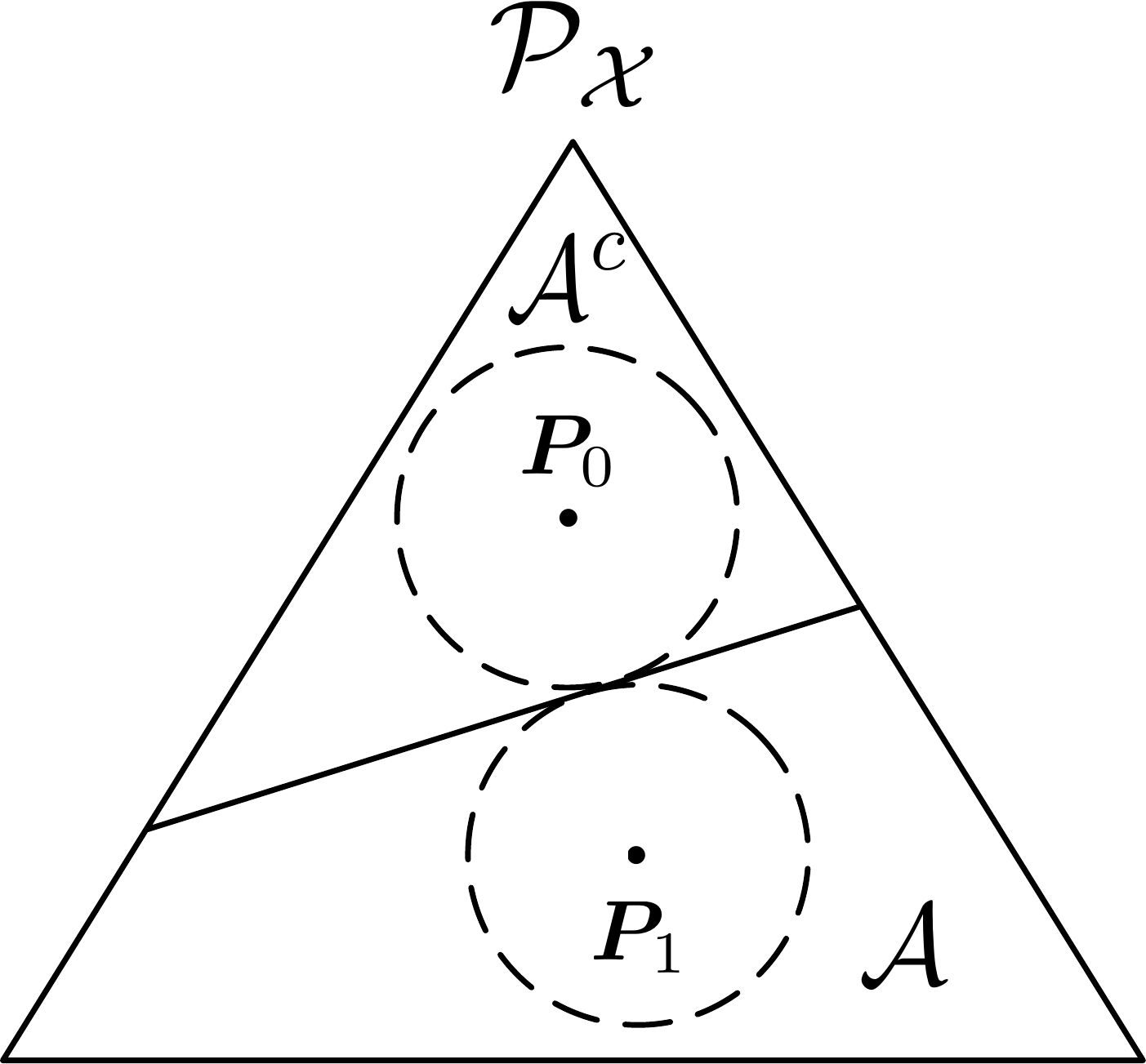} %
		\caption{Relation between $\mcal{A},\,\mcal{A}^c$ and $B_{r^*}(\bm{P}_0),\,B_{r^*}(\bm{P}_1)$ }%
		\label{fig:sanov}
\end{figure}

It is straightforward to see that  $\mcal{A}^c$ contains $B_{r^*}(\bm{P}_0)$ and $\mcal{A}$ contains $B_{r^*}(\bm{P}_1)$, since we must have
\begin{enumerate}
	\item $\forall\,  T \in B_{r^*}(\bm{P}_0)$, $f_{\bm{P}_0}(T) < f_{\bm{P}_1}(T)$,
	\item $\forall\,  T \in B_{r^*}(\bm{P}_1)$, $f_{\bm{P}_0}(T) > f_{\bm{P}_1}(T)$.
\end{enumerate}
Otherwise $B_{r^*}(\bm{P}_0)$ intersects $B_{r^*}(\bm{P}_1)$, violating our assumption on $r^*$. Also notice that $\mcal{A}$, $\mcal{A}^c$ are disjoint, so 
$$ \mcal{A}^c \cap B_{r^*}(\bm{P}_0) = \mcal{A} \cap B_{r^*}(\bm{P}_1) = \emptyset,$$
implying that 
$$ \mcal{A}^c \subseteq B_{r^*}(\bm{P}_1)^c, \, \mcal{A} \subseteq B_{r^*}(\bm{P}_0)^c. $$
Therefore, we have 
\begin{equation*}
	\begin{cases}
		\inf_{T\in\mcal{A}} f_{\bm{P}_0}(T) \geq \inf_{T\in B_{r^*}(\bm{P}_0)^c} f_{\bm{P}_0}(T) \geq r^* \\
		\inf_{T\in\mcal{A}^c} f_{\bm{P}_1}(T)\geq \inf_{T\in B_{r^*}(\bm{P}_1)^c} f_{\bm{P}_1}(T) \geq r^*,
	\end{cases}
\end{equation*}
proving the achievability part.
\end{proofpart}

\begin{proofpart}[Converse]

We show that for any test $\phi^{(n)}$, the exponent of the average probability of error greater than $r^*$ leads to contradiction. Suppose the type-I and type-II error exponents of $\phi^{(n)}$ are $r_1, r_2$ respectively, and $r_1> r^*$, $r_2>r^*$. By Lemma~\ref{lemma_2}, we only need to consider symmetric tests, that is, tests depend only on the type. Therefore, we can write the acceptance region of $\mcal{H}_0$, $\mcal{H}_1$ as 
\begin{equation*}
	\begin{cases}
		\mcal{B}_1^{(n)} = \lbp \Pi_{x^n} : \phi^{(n)}(x^n) = 1\rbp\\
		\mcal{B}_0^{(n)} = \lbp \Pi_{x^n} : \phi^{(n)}(x^n) = 0\rbp
	.\end{cases}
\end{equation*}
The exponents of type-I and type-II errors thus are greater then $r_1, r_2$ respectively, we have
\begin{equation}\label{eq1}
	\begin{cases}
		 \liminf\limits_{n\ra\infty} \lbp\min\limits_{T\in\mcal{B}_1^{(n)}} f_{\bm{P}_0}(T)\rbp = r_1 > r^*\\
		 \liminf\limits_{n\ra\infty} \lbp\min\limits_{T\in\mcal{B}_0^{(n)}} f_{\bm{P}_1}(T)\rbp = r_2 > r^*
	.\end{cases}
\end{equation}
Define $\min\lbp r_1, r_2 \rbp = \tilde{r}$, and $\delta \eqDef \lp\tilde{r} - r^*\rp/2 >0$.
By \eqref{eq1}, there exists $M$ large enough, such that for all $n > M$,
\begin{equation*}
	\begin{cases}
		 \min\limits_{T\in\mcal{B}_1^{(n)}} f_{\bm{P}_0}(T) > \tilde{r} - \delta > r^*\\
		 \min\limits_{T\in\mcal{B}_0^{(n)}} f_{\bm{P}_1}(T) > \tilde{r} - \delta > r^*
	.\end{cases}
\end{equation*}
We further define
\begin{equation*}
	\begin{cases}
		 \mcal{B}_1 = \bigcup\limits_{n>M} \mcal{B}_1^{(n)}\\
		 \mcal{B}_0 = \bigcup\limits_{n>M} \mcal{B}_0^{(n)}
	.\end{cases}
\end{equation*}
We see that
\begin{enumerate}
	\item $\mcal{B}_0 \cup \mcal{B}_1$ are dense in $\mcal{P_X}$, since
	$$\mcal{B}_0^{(n)} \cup \mcal{B}_1^{(n)} = \mcal{P}_n,$$
	and $\bigcup\limits_{n>M}\mcal{P}_n$ is dense in $\mcal{P_X}$.
	So we have 
	\begin{equation}\label{eq:chrff_cv_1}
	\lp\text{cl }\mcal{B}_0 \cup \text{cl }\mcal{B}_1\rp^{c} = \lp\text{cl }\mcal{B}_0\rp^{c} \cap \lp\text{cl }\mcal{B}_1\rp^{c} = \emptyset.  
	\end{equation}
	\item By construction, 
		\begin{equation}\label{eq:chrff_cv_2}
			\begin{cases}
			 \inf\limits_{T\in\mcal{B}_1} f_{\bm{P}_0}(T) = \min\limits_{T\in\text{cl }\mcal{B}_1} f_{\bm{P}_0}(T) > \tilde{r} - \delta > r^* \\
			 \inf\limits_{T\in\mcal{B}_0} f_{\bm{P}_1}(T) = \min\limits_{T\in\text{cl }\mcal{B}_0} f_{\bm{P}_1}(T) > \tilde{r} - \delta > r^*
			.\end{cases}
		\end{equation}
\end{enumerate}
From \eqref{eq:chrff_cv_2}, we have 
\begin{equation*}
	\begin{cases}
		B_{(\tilde{r}-\delta)}(\bm{P}_0) \subseteq \lp\text{cl } \mcal{B}_1\rp^c \nonumber\\
		B_{(\tilde{r}-\delta)}(\bm{P}_1) \subseteq \lp\text{cl } \mcal{B}_0\rp^c,
	\end{cases}
\end{equation*}
and by \eqref{eq:chrff_cv_1} $B_{(\tilde{r}-\delta)}(\bm{P}_0) \cap B_{(\tilde{r}-\delta)}(\bm{P}_1) = \emptyset$. However, this violates our assumption that $r^*$ is the supreme of radius such that the two sets do not overlap. This proves the converse part.
\end{proofpart}
\end{IEEEproof}
\begin{remark}
	In Theorem~\ref{thm:eff_test}, we provide an asymptotically optimal test based on an information-geometric perspective. However, we do not specify the exact error exponent. As stated in the proof, the optimal exponent of average probability of error can be obtain by solving the information projection problem:
	$$ \min_{T\in\mcal{A}} f_{\bm{P}_0}(T), $$
	where $\mcal{A}$ is the acceptance region of $\phi_\text{eff}$. The optimization problem, though convex, is hard to obtain a closed-form expression, but we can still evaluate it numerically.
\end{remark}

\subsection{Characterization of Achievable Exponent Region $\mcal{R}$} 
One can generalize the result from Theorem~\ref{thm:eff_test}. Define the following test:
$$ \phi_\lambda(x^n) \eqDef \mathbbm{1}_{\lbp f_{\bm{P}_0}(\Pi_{x^n})- f_{\bm{P}_1}(\Pi_{x^n}) \geq \lambda \rbp}, $$
where $\lambda\in\lb -f_{\bm{P}_1}\lp M_0(\bm{\alpha})\rp, f_{\bm{P}_0}\lp M_1(\bm{\alpha})\rp \rb$. Following a similar idea in the proof of Theorem~\ref{thm:eff_test}, one can show that $\phi_\lambda$ is optimal in a sense that for any test $\phi$ and $\forall\,\lambda$, 
$$ E_0(\phi) \geq E_0(\phi_\lambda) \Rightarrow  E_1(\phi) \leq E_1(\phi_\lambda),$$ and
$$ E_1(\phi) \geq E_1(\phi_\lambda) \Rightarrow  E_0(\phi) \leq E_0(\phi_\lambda),$$
where
$\lp E_0(\phi), E_1(\phi) \rp$ are the error exponents with respect to test $\phi$ : 
\begin{equation*}
	\begin{cases}
		E_0(\phi) \eqDef \liminf\limits_{n\ra\infty}\lbp -\frac{1}{n}\log \mathsf{P_F}^{(n)}(\phi)\rbp\\
		E_1(\phi) \eqDef \liminf\limits_{n\ra\infty}\lbp -\frac{1}{n}\log \mathsf{P_M}^{(n)}(\phi)\rbp.
	\end{cases}
\end{equation*}
To obtain a parametrization of the boundary of $\mcal{R}$, it suffices to solve the following information projection problem:
\begin{equation*}
	\begin{cases}
		E_0(\lambda) \eqDef \inf_{T\in\mcal{A}_\lambda} f_{\bm{P}_0}(T)\\
		E_1(\lambda) \eqDef \inf_{T\in\lp\mcal{A}_\lambda\rp^c} f_{\bm{P}_1}(T),
	\end{cases}
\end{equation*}
where $\mcal{A}_\lambda \eqDef \lbp f_{\bm{P}_0}(\Pi_{x^n})- f_{\bm{P}_1}(\Pi_{x^n}) \geq \lambda \rbp$ is the acceptance region of $\phi_\lambda$. Therefore, $\lp E_0(\lambda), E_1(\lambda) \rp$ parametrizes the boundary of $\mcal{R}$, for $\lambda\in\lb -f_{\bm{P}_1}\lp M_0(\bm{\alpha})\rp, f_{\bm{P}_0}\lp M_1(\bm{\alpha})\rp \rb$. 

\begin{figure}[htbp]
		\centering
		\includegraphics[width=0.4\linewidth]{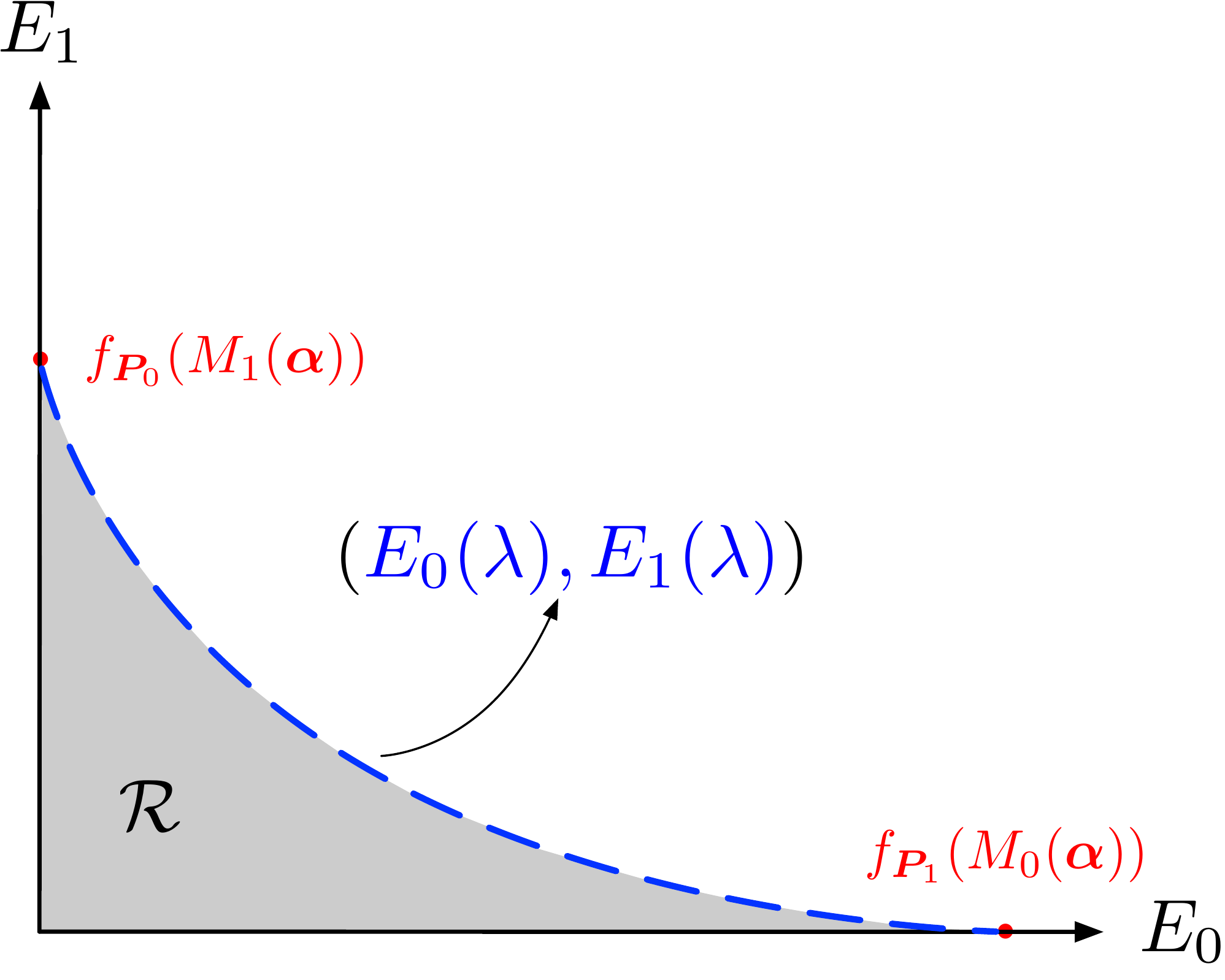} %
		\caption{Illustration of $\lp E_0(\lambda), E_1(\lambda) \rp$ }%
		\label{fig:ach}
\end{figure}
In particular, we see that for the corners $\lambda = f_{\bm{P}_0}\lp M_1(\bm{\alpha})\rp$ and $\lambda = -f_{\bm{P}_1}\lp M_0(\bm{\alpha})\rp$, we obtain the same results as in Neyman-Pearson regime (Theorem~\ref{thm:asymptotics}).
Note that although the information-projection problem is a convex optimization problem, the closed-form expression remains unknown.

%% file: sec_discussion.tex
\subsection{Extension to Polish $\mcal{X}$}
Theorem~\ref{thm:optimal_test} characterizes the optimal test in the anonymous detection problem, where only a few conditions on the $\sigma$-field $\mcal{F}$ are required. In Theorem~\ref{thm:asymptotics}, we further assume the alphabet $\mcal{X}$ is finite, in order to apply large deviation tools based on the method of types (see Remark~\ref{rmk:asymptotic} for discussion). However, the the optimal exponent of the type-II error probability, given by the result of Theorem~\ref{thm:asymptotics}, depends only on the possible distributions under $\mcal{H}_\theta$, and hence it is interesting to see if one can remove the assumption that $\mcal{X}$ being finite. Recall that in the proof, the main tool we employed is the generalized version of Sanov's theorem (see Lemma~\ref{lemma:general_sanov}), and thus the question turns out to be whether it is possible to prove Lemma~\ref{lemma:general_sanov} without using method of types. Surprisingly, the answer is yes if $\mcal{X}$ is a Polish space (a completely separable metrizabla topological space). If $\mcal{X}$ is Polish, the space of all probability measures on $\mcal{X}$ ($\mcal{P_X}$) is also Polish, equipped with weak-topology induced by weak convergence. One can choose, for example, Levy-Prokhorov metric on $\mcal{P_X}$. The proof of standard Sanov's Theorem on Polish $\mcal{X}$, however, is far more complicated than the case of finite $\mcal{X}$, see \cite{Hol00,DemZei10} for detailed proof. Lemma~\ref{lemma:general_sanov} for Polish $\mcal{X}$ can be proved with similar techniques. Nevertheless, in order not to digress further from the subject, we only present a proof for finite $\mcal{X}$ in this paper.

\subsection{The Benefit of Partial Information about the Group Assignment}

From Figure~\ref{fig:anony}, we see that in some cases, the type-II error exponent can be pushed to zero, making reliable detection no longer possible. If each sensor is allowed to transmit a few bits of information to \emph{partially reveal} their groups, how such partial information can improve the type-II error exponent? Formally speaking, we assume that the total number of groups is $K$, and each sensor can transmit $L$ bits (with $L < \log K$) through a noiseless channel to the fusion center, providing partial information about the group that it belongs to. 

Unsurprisingly, the optimal strategy is the \emph{cluster-and-detect} approach, that is, we first \emph{cluster} the $K$ groups into
$2^L$ super-groups, and each sensor sends $L$ bits to indicate which \emph{super-groups} it belongs to. Inside each super-group, we adopt the optimal anonymous hypothesis testing, and between super-groups, the problem boils down to the equivalent informed hypothesis testing, and hence standard likelihood ratio test can be applied there.

However, the difficulty lies in the clustering step: even the fusion center knows the distribution of each group, the optimal clustering algorithm is indeed a discrete optimization problem and thus NP-hard. When the group number $K$ is large enough, it is intractable to find the optimal clustering. 
Nevertheless, some suboptimal algorithms suggested by heuristic do demonstrate that this partial information can significantly ameliorate the performance loss caused by anonymity. Below is a numerical example, showing the benefit of partial information.

\begin{figure}[htbp]
		\centering
		\includegraphics[width=0.6\linewidth]{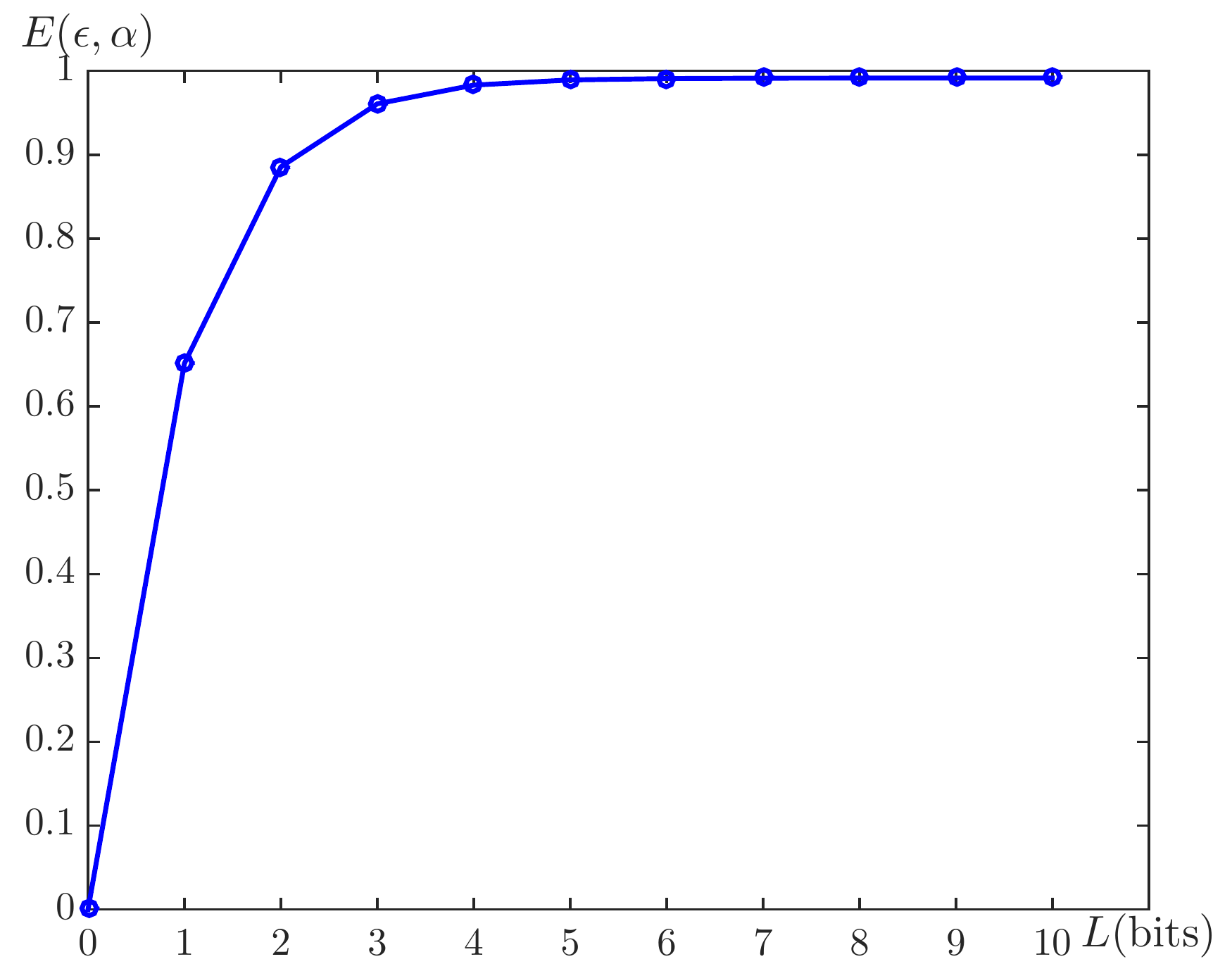} %
		\caption{Exponents with Partial Information }%
		\label{fig:e_partial}
\end{figure}

In the example, we assume their are totally $K=1024$ ($2^{10}$) groups, and each group accounts for $1/K$ proportion of total sensors, that is, $\bm{\alpha} = [\frac{1}{K},...,\frac{1}{K}]^\intercal$. For the sensors in the $k$-th group, their observations follow i.i.d. distribution $\Ber(\theta_k)$ under $\mcal{H}_0$, and follow i.i.d. $\Ber(1-\theta_k)$ under $\mcal{H}_1$, with $\theta_k = \frac{k}{K}, k=1,...,K$. Suppose there are $L$ bits available for each sensor to partially inform the fusion center the group it belongs to, then as the clustering-detection algorithm suggests, we first cluster the $K$ groups into $2^L$ super-groups and then apply anonymous hypothesis testing inside each super-group. As the numerical evaluation in Figure~\ref{fig:e_partial} illustrates, even with few bits, say, $L=1 \text{ or }2$, type-II error exponents are significantly improved.

%% file: sec_conclusion.tex
In this paper, we explore the heterogeneous distributed detection problem with sensor anonymity. To address sensor anonymity, a composite hypothesis testing approach is taken. Focusing on the Neyman-Pearson setting, we provide an optimal test, and characterize the exponent of type-II error probability for the case that $\mcal{X}$ is finite. Unlike the settings considered in robust hypothesis testing literatures \cite{Huber_65,HuberStrassen_73,VeeravalliBasar_94}, since the hypothesis classes considered in our framework are discrete, the least favorable distribution might not exist. To circumvent the difficulty, we map the original problem into an auxiliary space by employing the symmetric property of the hypothesis classes, in which the original composite hypothesis testing problem becomes a simple hypothesis testing problem. Therefore, Neyman-Pearson lemma can be applied to obtain an optimal test, which is a randomized threshold test based on the ratio of the uniform mixture of all the possible distributions under $\mcal{H}_0$ to  the uniform mixture of those under $\mcal{H}_1$. For the asymptotic regime, we analyze the type-II error exponent using method of types and show that the optimal exponent is the minimization of linear combination of KL-divergences, with the $k$-th term being $\KLD{U_k}{P_{1;k}}$ and $\alpha_k$ being the coefficient, for $k=1,...,K$. The minimization is over all possible distributions $U_1,...,U_K$ such that $\sum_{k=1}^{K}\alpha_kU_k = \sum_{k=1}^{K}\alpha_kP_{0;k}$. We further extend our result to Chernoff's regime, and indicate that the exponent region can be obtained by solving a convex optimization problem.

There are still many open problems in anonymous heterogeneous hypothesis testing. For example, the closed-form expression for the exponents in asymptotic regime, even in Neyman-Pearson formulation, are still unknown. Besides, the solution of information projection is conjectured to have similar form like tilted-distributions, as the classical results in simple hypothesis testing suggested. In addition to hypothesis testing, it is also interesting to investigate other problems such as regression, estimation, or pattern recognition under the anonymous setting.

%% file: appendix_prop.tex
\setcounter{proofpart}{0}
\begin{IEEEproof}[proof of Proposition~\ref{prop:cvx}]
	Since the optimal type-II exponent does not depend on $\epsilon$, we denote it  as $E^*(\bm{\alpha})$ and for simplicity. It suffices to show
	$$ E^*(\lambda\bm{\alpha}_1+(1-\lambda)\bm{\alpha}_2) \leq \lambda E^*(\bm{\alpha}_1)+(1-\lambda)E^*(\bm{\alpha}_2), \, \forall \lambda\in[0,1].$$
	First, let 
	\begin{align*}
		E^*(\bm{\alpha}_1) &= \sum_{k=1}^K\alpha_{1k}\KLD{U^*_{1k}}{P_{1;k}}\\
		E^*(\bm{\alpha}_2) &= \sum_{k=1}^K\alpha_{2k}\KLD{U^*_{2k}}{P_{1;k}}
	\end{align*}
	where $\bm{\alpha}_1 = [\alpha_{11},...,\alpha_{1K}]^\intercal,\bm{\alpha}_2 = [\alpha_{21},...,\alpha_{2K}]^\intercal$, and $\bm{U}_1^*\eqDef [U^*_{11},...,U^*_{1K}],\bm{U}_2^*\eqDef [U^*_{21},...,U^*_{2K}]$ are the minimizers of \eqref{eq:lowerbd}. Then, by the convexity of KL divergence, we have
	\begin{align}
		\lambda E^*(\bm{\alpha}_1)+(1-\lambda)E^*(\bm{\alpha}_2) 
		= &\sum_{k=1}^K\lambda\alpha_{1k}\KLD{U^*_{1k}}{P_{1;k}}+(1-\lambda)\alpha_{2k}\KLD{U^*_{2k}}{P_{1;k}} \nonumber \\
		\geq &\sum_{k=1}^K (\lambda\alpha_{1k}+(1-\lambda)\alpha_{2k})\KLD{ \frac{\lambda\alpha_{1k}U^*_{1k}+(1-\lambda)\alpha_{2k}U^*_{2k}}{\lambda\alpha_{1k}+(1-\lambda)\alpha_{2k}} }{P_{1;k}} \label{eq:cvx_ach}
	\end{align}
	Now we claim that $\tilde{\bm{U}}\eqDef \lp\frac{\lambda\alpha_{1k}U^*_{1k}+(1-\lambda)\alpha_{2k}U^*_{2k}}{\lambda\alpha_{1k}+(1-\lambda)\alpha_{2k}}\rp_{k=1,...,K}$ satisfies 
	\begin{equation}\label{eq:ach_constraint_1}
		(\lambda\bm{\alpha}_1+(1-\lambda)\bm{\alpha}_2)^\intercal\tilde{\bm{U}}=(\lambda\bm{\alpha}_1+(1-\lambda)\bm{\alpha}_2)^\intercal\bm{P}_0,
	\end{equation} 
	and thus
	\begin{align*}
		\eqref{eq:cvx_ach} = &\sum\nolimits_{k=1}^K(\lambda\alpha_{1k}+(1-\lambda)\alpha_{2k})\KLD{\tilde{U}_k}{P_{1;k}}\\
		\geq &\min\limits_{\substack{\bm{U}\in(\mcal{P_X})^K \\(\lambda\bm{\alpha}_1+(1-\lambda)\bm{\alpha}_2)^\intercal\bm{U}=(\lambda\bm{\alpha}_1+(1-\lambda)\bm{\alpha}_2)^\intercal\bm{P}_0}
		}\, 
		\sum\nolimits_{k=1}^K(\lambda\alpha_{1k}+(1-\lambda)\alpha_{2k})\KLD{U_k}{P_{1;k}} \\
		= &E^*(\lambda\bm{\alpha}_1+(1-\lambda)\bm{\alpha}_2).
	\end{align*}	
	To show \eqref{eq:ach_constraint_1}, we notice that $\bm{U}^*_1$, $\bm{U}^*_2$ satisfy the constraints 
	\begin{equation}\label{eq:ach_constraint_2}
	\bm{\alpha}_1^\intercal\bm{U}^*_1 = \bm{\alpha}_1^\intercal\bm{P}_0, \,\bm{\alpha}_2^\intercal\bm{U}^*_2 = \bm{\alpha}_2^\intercal\bm{P}_0.
	\end{equation}
	Then we have 
	\begin{align*}
		&(\lambda\bm{\alpha}_1+(1-\lambda)\bm{\alpha}_2)^\intercal\tilde{\bm{U}}\\
		=& \sum_{k=1}^K (\lambda\alpha_{1k}+(1-\lambda)\alpha_{2k})\lp\frac{\lambda\alpha_{1k}U^*_{1k}+(1-\lambda)\alpha_{2k}U^*_{2k}}{\lambda\alpha_{1k}+(1-\lambda)\alpha_{2k}}\rp\\
		=&\sum_{k=1}^K\lambda\alpha_{1k}U^*_{1k}+(1-\lambda)\alpha_{2k}U^*_{2k}\\
		=&\lambda\bm{\alpha}_1^\intercal\bm{U}^*_1+(1-\lambda)\bm{\alpha}_2^\intercal\bm{U}^*_2\\
		=&(\lambda\bm{\alpha}_1+(1-\lambda)\bm{\alpha}_2)^\intercal\bm{P}_0,
	\end{align*}
	which completes the proof.
\end{IEEEproof}

%% file: appendix_general_sanov.tex
\begin{IEEEproof}[proof of Lemma~\ref{lemma:general_sanov}]
	First, observe that 
	since $\text{int }\Gamma$ is open, the set 
	$$ \tilde{\Gamma} \eqDef \lbp (U_1,...,U_K) \mid \bm{\alpha}^\intercal\bm{U}\in \text{int } \Gamma \rbp \subset \lp\mcal{P_X}\rp^K$$
	is open too. This is because the mapping $g(\bm{U}) = \bm{\alpha}^\intercal\bm{U}$ is continuous, so the pre-image preserves the openness (under standard topology). Therefore, we can find a sequence 
	$$ \lbp \bm{U}^{(n)} \in \lp\mcal{P}_{n_1}\times\cdots\times\mcal{P}_{n_K}\rp \cap \tilde{\Gamma}\rbp,$$ such that  
	$$ \sum_k \alpha_k\KLD{U_k^{(n)}}{P_{\theta;k}} \ra -\inf_{\substack{ (U_1,...,U_K)\in\lp\mcal{P_X}\rp^K \\ \bm{\alpha}^\intercal\bm{U}\in \text{int } \Gamma}} \sum_k \alpha_k\KLD{U_k}{P_{\theta;k}}, $$
	where the limit is taken such that $\frac{n_k}{n}\ra\alpha_k$.
	So we have 
	\begin{align*}
		\mbb{P}_{\theta;\sigma}\lbp \Pi_{x^n} \in \Gamma \rbp 
	&= \sum\limits_{\substack{ (U_1,...,U_K)\in\mcal{P}_{n_1}\times\cdots\times\mcal{P}_{n_K} \\ \bm{\alpha}^\intercal\bm{U}\in \Gamma}} \prod_{k=1}^K P_{\theta;k}^{\otimes n_k}\lbp T_{n_k}(U_k) \rbp \\
	&\geq \sum\limits_{\substack{ (U_1,...,U_K)\in\mcal{P}_{n_1}\times\cdots\times\mcal{P}_{n_K} \\ \bm{\alpha}^\intercal\bm{U}\in \text{int } \Gamma}} \prod_{k=1}^K P_{\theta;k}^{\otimes n_k}\lbp T_{n_k}(U_k) \rbp\\
	&\geq \max_{\substack{ (U_1,...,U_K)\in\mcal{P}_{n_1}\times\cdots\times\mcal{P}_{n_K} \\ \bm{\alpha}^\intercal\bm{U}\in \text{int } \Gamma}}\prod_{k=1}^K P_{\theta;k}^{\otimes n_k}\lbp T_{n_k}\lp U_k^{(n)}\rp \rbp\\
	&\overset{(a)}{\geq} \max_{\substack{ (U_1,...,U_K)\in\mcal{P}_{n_1}\times\cdots\times\mcal{P}_{n_K} \\ \bm{\alpha}^\intercal\bm{U}\in \text{int } \Gamma}}\lp\frac{1}{(n_k+1)^{\lba\mcal{X}\rba}}\rp2^{\sum_{k=1}^K n_k\KLD{U_k^{(n)}}{P_{\theta;k}}},
	\end{align*}
	where inequality (a) holds by Lemma~\ref{lemma:type}.
	Thus we have 
	$$ \frac{1}{n}\log \mbb{P}_{\theta;\sigma}\lbp \Pi_{x^n} \in \Gamma \rbp \geq
	- \min_{\substack{ (U_1,...,U_K)\in\mcal{P}_{n_1}\times\cdots\times\mcal{P}_{n_K} \\ \bm{\alpha}^\intercal\bm{U}\in \text{int } \Gamma}} \lp \sum_{k=1}^K \frac{n_k}{n}\KLD{U_k^{(n)}}{P_{\theta;k}} + o(1)\rp.$$
	As $n\ra\infty$ such that $\frac{n_k}{n}\ra \alpha_k$, we see that  
	$$ -\inf_{\substack{ (U_1,...,U_K)\in\lp\mcal{P_X}\rp^K \\ \bm{\alpha}^\intercal\bm{U}\in\text{int }\Gamma}} \sum_k \alpha_k\KLD{U_k}{P_{\theta;k}} \leq \liminf_{n\ra\infty} \frac{1}{n}\log \mbb{P}_{\theta;\sigma}\lbp \Pi_{x^n} \in \Gamma \rbp. $$

 	On the other hand, for the upper bound, consider 
 	\begin{align*}
		\mbb{P}_{\theta;\sigma}\lbp \Pi_{x^n} \in \Gamma \rbp 
	&= \sum\limits_{\substack{ (U_1,...,U_K)\in\mcal{P}_{n_1}\times\cdots\times\mcal{P}_{n_K} \\ \bm{\alpha}^\intercal\bm{U}\in \Gamma}} \prod_{k=1}^K P_{\theta;k}^{\otimes n_k}\lbp T_{n_k}(U_k) \rbp \\
	&\overset{(a)}{\leq} \sum\limits_{\substack{ (U_1,...,U_K)\in\mcal{P}_{n_1}\times\cdots\times\mcal{P}_{n_K} \\ \bm{\alpha}^\intercal\bm{U}\in \Gamma}} 2^{\sum_{k=1}^K \KLD{U_k^{(n)}}{P_{\theta;k}}}\\
	&\leq \lp\prod_k\lba \mcal{P}_{n^k} \rba \rp 2^{\sum_{k=1}^K n_k \KLD{U_k^{(n)}}{P_{\theta;k}}}\\
	& \overset{(b)}{=} 2^{ \lp \sum_{k=1}^K n_k\KLD{U_k^{(n)}}{P_{\theta;k}} + o(1)\rp},
	\end{align*}
	where where inequality (a) holds by Lemma~\ref{lemma:type}, and (b) holds due to the cardinality bound Lemma~\ref{lemma:Pn_bound}.

	As $n\ra\infty$ and $\frac{n_k}{n}\ra\alpha_k$, we have 
	$$  \limsup_{n\ra\infty} \frac{1}{n}\log \mbb{P}_{\theta;\sigma}\lbp \Pi_{x^n} \in \Gamma \rbp 
 		\leq  -\inf_{\substack{ (U_1,...,U_K)\in\lp\mcal{P_X}\rp^K \\ \bm{\alpha}^\intercal\bm{U}\in\Gamma}} \sum_k \alpha_k\KLD{U_k}{P_{\theta;k}}.$$
 	Notice that for the case $\mcal{X}$ finite, the infimum takes over $\Gamma$ is equal to that one takes in the closure of $\Gamma$, since we can use standard topology to find a sequence approaching to the limit point. Thus the proof is complete.
\end{IEEEproof}

%% file: appendix_property_d.tex
\begin{IEEEproof}[proof of Lemma~\ref{lemma:property_d}]
\setcounter{proofpart}{0}
Let $\bm{Q}\in\lp\mcal{P_X}\rp^K$ be a K-tuple of probability measure on $\mcal{X}$. We first show that 
$$\mcal{C}_{\bm{Q}} \eqDef \lbp T\in\mcal{P_X} : f_{\bm{Q}}(T) < \infty \rbp$$ 
is a compact set.
\begin{proofpart}[Compactness]
	Observe that $f_{\bm{Q}}(T) < \infty$ if and only if there exists a $\bm{P} = (P_1,...,P_K)\in \lp\mcal{P_X}\rp^K$, such that 
	\begin{enumerate}
		\item $\bm{\alpha}^\intercal\bm{P} = T$
		\item for all $i=1,...,K$, $P_i\ll Q_i$.
	\end{enumerate}
	Therefore, let us denote
	$$ \mcal{M}_{\bm{Q}} \eqDef\lbp \bm{P}\in \lp\mcal{P_X}\rp^K : P_i\ll Q_i, \, \forall i = 1,...,K\rbp\subseteq\lp\mcal{P_X}\rp^K.$$
	We claim that $\mcal{M}_{\bm{Q}}$ is a compact set, and thus 
	$$ \mcal{C}_{\bm{Q}} = \lbp \bm{\alpha}^\intercal\bm{P} \mid \bm{P} \in \mcal{M}_{\bm{Q}} \rbp$$
	is also compact, since $\bm{\alpha}^\intercal\bm{P}$ is a linear mapping from $\lp\mcal{P_X}\rp^K$ to $\mcal{P_X}$ so compactness is preserved.
	To prove the claim, it suffices to show that $\mcal{M}_{\bm{Q}}$ is a closed set, because the boundness is directly followed by the boundness of $\lp\mcal{P_X}\rp^K$. It is equivalent to show 
	$$  \mcal{M}_{\bm{Q}}^C =  
	\lbp \bm{P}\in \lp\mcal{P_X}\rp^K : P_i \not\ll Q_i, \, \text{for some } i \rbp$$
	is open. 
	Notice that 
	$$ \lbp \bm{P}\in \lp\mcal{P_X}\rp^K : P_i \not\ll Q_i, \, \text{for some } i \rbp = 
	\bigcup_{i=1}^K \lbp \bm{P}\in \lp\mcal{P_X}\rp^K : P_i \not\ll Q_i \rbp,$$
	so it suffices to show $\lbp \bm{P}\in \lp\mcal{P_X}\rp^K : P_i \not\ll Q_i \rbp$ is open for all $i$.
	Assume $P_i \not\ll Q_i$. Then there must exist some measurable event $\mcal{E}\subset \mcal{X}$, such that $Q_i(\mcal{E}) = 0$, and $P_i(\mcal{E})=\epsilon >0$. Therefore, if $\mcal{X}$ is finite and thus $\mcal{P_X}$ equipped with total-variation distance (i.e. one norm), then obviously for any $\tilde{Q}$ such that $\lVert \tilde{Q} - P_i \rVert < \frac{\epsilon}{2}$, $\tilde{Q}\not\ll Q_i$. Hence $\mcal{M}_{\bm{Q}}^C$ is open, proving the claim.
	\begin{remark}
	If $\mcal{X}$ is Polish, then $\mcal{P_X}$ is equipped with Prokhorov's metric, and one can use similar argument to show that $\mcal{M}_{\bm{Q}}^C$ is open.
	\end{remark}
\end{proofpart}
Next, we show that $f_{\bm{Q}}(\cdot)$ is a convex function, so the convexity of $\mcal{C}_{\bm{Q}}$ follows: for all $T_1,T_2 \in \mcal{C}_{\bm{Q}}$,
\begin{equation}\label{eq:cvx_d}
f_{\bm{Q}}(\lambda T_1+(1-\lambda)T_2)\leq \lambda f_{\bm{Q}}(T_1)+(1-\lambda)f_{\bm{Q}}(T_2) < \infty,
\end{equation}
implying $\lambda T_1+(1-\lambda)T_2 \in \mcal{C}_{\bm{Q}}$.
\begin{proofpart}[Convexity]
To show \eqref{eq:cvx_d}, we observe
\begin{align*}
	&\lambda f_{\bm{Q}}(T_1)+(1-\lambda)f_{\bm{Q}}(T_2) \\
= & \inf_{\bm{U}:\bm{\alpha}^\intercal\bm{U} = T_1} \lambda\sum_k \alpha_k \KLD{U_k}{P_k} +\inf_{\bm{V}:\bm{\alpha}^\intercal\bm{V}=T_2} (1-\lambda)\sum_k \alpha_k \KLD{V_k}{P_k}\\
\overset{(a)}{\geq} & \inf_{\bm{U},\bm{V}: \bm{\alpha}^\intercal\bm{U} = T_1, \bm{\alpha}^\intercal\bm{U} = T_2}
\sum_k \alpha_k\KLD{\lambda U_k +(1-\lambda) V_k}{P_k}\\
\overset{(b)}{\geq} & \inf_{\bm{P}:\bm{\alpha}^\intercal\bm{P} = \lambda T_1+(1-\lambda)T_2} \sum_k \alpha_k\KLD{Q_k}{P_k}\\
= &f_{\bm{Q}}(\lambda T_1+(1-\lambda)T_2),
\end{align*}
where (a) is due to the convexity of KL-divergence, and (b) is because

$$ \bm{\alpha}^\intercal\bm{U} = T_1, \bm{\alpha}^\intercal\bm{V} = T_2 \Rightarrow  \bm{\alpha}^\intercal\lp\lambda \bm{U} +(1-\lambda) \bm{V} \rp = \lambda T_1+(1-\lambda)T_2.$$
Therefore, we conclude that $f_{\bm{Q}}(\cdot)$ is a convex function and $\mcal{C}_{\bm{Q}}$ is a convex set.
\end{proofpart}
At the final step, we show $f_{\bm{Q}}(\cdot)$ is a continuous function on $\mcal{C}_{\bm{Q}}$. Notice that the convexity of $f_{\bm{Q}}(\cdot)$ only guarantees the continuity on the interior of $\mcal{C}_{\bm{Q}}$, and thus we need to additionally check the boundary points.
\begin{remark}
	Note that in general, the interior of $\mcal{C}_{\bm{Q}}$ may be an empty set since it may lie in a subspace of $\mcal{P_X}$. Alternatively, we can define a point $\bm{P}$ being interior, if it can be written as 
	$$\lambda\bm{U}+(1-\lambda)\bm{V}, \text{ for some }\lambda \in (0,1) \text{, and some } \bm{V},\bm{U} \in \mcal{C}_{\bm{Q}}.$$
\end{remark}
\begin{proofpart}[Continuity]
First, if the interior of $\mcal{C}_{\bm{Q}}$ is empty, then by the convexity, either $\mcal{C}_{\bm{Q}}$ is a empty set, or it is a singleton. For both cases, the continuity holds obviously. Hence without losing of generality, we assume that the interior of $\mcal{C}_{\bm{Q}}$ is non-empty, and $T_0$ is an interior point.

Then for any $T \in \mcal{C}_{\bm{Q}}$, we can construct a sequence 
$T_n\in \mcal{C}_{\bm{Q}}$, $T_n \ra T$. For example, one can let $T_n = \lambda_n T_0+(1-\lambda_n)T$, with $\lambda_n\ra 0 $.
Let $\bm{U}^{(n)} = (U^{(n)}_1,...,U^{(n)}_K) \in \lp\mcal{P_X}\rp^K$ be a sequence such that 
\begin{enumerate}
 	\item $\bm{\alpha}^\intercal \bm{U}^{(n)} = T_n$
 	\item $\bm{U}^{(n)}$ achieves the infimum of $f_{\bm{Q}}(T_n)$ :
 		$$ \sum_{k=1}^K \alpha_k\KLD{U^{(n)}_k}{P_k} = \inf_{\bm{V} : \bm{\alpha}^\intercal\bm{V} = T_n} \sum_{k=1}^K \alpha_k\KLD{V_k}{P_k} = f_{\bm{Q}}(T_n). $$
 	Notice that the infimum can always be achieved since $g(\bm{V}) \eqDef\sum_{k=1}^K \alpha_k\KLD{V_k}{P_k}$ is a continuous function over the compact set $\mcal{M}_{\bm{Q}}$.
 \end{enumerate} 
By construction, $\bm{U}^{(n)}$ is a sequence in a compact set $\mcal{M}_{\bm{Q}}$, and hence by Bolzano-Weierstrass theorem (see Chapter~1 in \cite{RoyFit10}, for example), there exists a convergent subsequence $\bm{U}^{(n_i)}$, and let us denote the convergent point 
$$ \lim_{i\ra\infty} \bm{U}^{(n_i)} = \bm{U}.$$
Since $\bm{\alpha}^\intercal\bm{U}^{(n_i)} = T_{n_i}$, and $T_{n_i} \ra T$, we have 
$$ \bm{\alpha}^\intercal\bm{U} = T. $$
Notice that the function $f(\bm{V}) \eqDef \sum_{k=1}^K \alpha_k\KLD{V_k}{P_k}$ is a continuous function over the compact set $\mcal{M}_{\bm{Q}}$, we must have 
$$ \lim_{n\ra\infty} f_{\bm{Q}}(T_n) = \lim_{i\ra\infty} f_{\bm{Q}}(T_{n_i}) = \sum_{k=1}^K \alpha_k\KLD{U_k}{P_k}, $$
and therefore
$$ f_{\bm{Q}}(T) = \inf_{\bm{U}:\bm{\alpha}^\intercal\bm{V} = T} \sum_{k=1}^K \alpha_k\KLD{V_k}{P_k} \leq \sum_{k=1}^K \alpha_k\KLD{U_k}{P_k} = \lim_{n\ra\infty} f_{\bm{Q}}(T_n).$$
On the other hand, by the convexity of $f_{\bm{Q}}(\cdot)$, we must have 
$$ f_{\bm{Q}}(T) \geq  f_{\bm{Q}}(T_n), \text{ for all } n \text{ large enough}.$$
Otherwise 
$$ f_{\bm{Q}}(\lambda T_0+ (1-\lambda) T) > \lambda f_{\bm{Q}}(T_0) + (1-\lambda)f_{\bm{Q}}(T), $$
for some $\lambda$ small enough, which violates the fact that $f_{\bm{Q}}(\cdot)$ is a convex function.
\end{proofpart}
\end{IEEEproof}

%% file: TIT_DDBA.bbl
\begin{thebibliography}{10}
\providecommand{\url}[1]{#1}
\csname url@samestyle\endcsname
\providecommand{\newblock}{\relax}
\providecommand{\bibinfo}[2]{#2}
\providecommand{\BIBentrySTDinterwordspacing}{\spaceskip=0pt\relax}
\providecommand{\BIBentryALTinterwordstretchfactor}{4}
\providecommand{\BIBentryALTinterwordspacing}{\spaceskip=\fontdimen2\font plus
\BIBentryALTinterwordstretchfactor\fontdimen3\font minus
  \fontdimen4\font\relax}
\providecommand{\BIBforeignlanguage}[2]{{%
\expandafter\ifx\csname l@#1\endcsname\relax
\typeout{** WARNING: IEEEtran.bst: No hyphenation pattern has been}%
\typeout{** loaded for the language `#1'. Using the pattern for}%
\typeout{** the default language instead.}%
\else
\language=\csname l@#1\endcsname
\fi
#2}}
\providecommand{\BIBdecl}{\relax}
\BIBdecl

\bibitem{Tsitsiklis_90}
J.~N. Tsitsiklis, ``Decentralized detection,'' in \emph{Advances in Statistical
  Signal Processing}, H.~V. Poor and J.~B. Thomas, Eds.\hskip 1em plus 0.5em
  minus 0.4em\relax JAI Press Inc., 1990, vol.~2.

\bibitem{Tsitsiklis_88}
------, ``Decentralized detection by a large number of sensors,''
  \emph{Mathematics of Control, Signals and Systems}, vol.~1, no.~2, pp.
  167--182, 1988.

\bibitem{MaranoMatta_09}
S.~Marano, V.~Matta, and L.~Tong, ``Distributed detection in the presence of
  {B}yzantine attacks,'' \emph{IEEE Transactions on Signal Processing},
  vol.~57, no.~1, pp. 16--29, January 2009.

\bibitem{Huber_65}
P.~J. Huber, ``A robust version of the probability ratio test,'' \emph{Annals
  of Mathematical Statistics}, vol.~36, no.~6, pp. 1753--1758, 1965.

\bibitem{HuberStrassen_73}
P.~J. Huber and V.~Strassen, ``Minimax tests and the {N}eyman-{P}earson lemma
  for capacities,'' \emph{Annals of Statistics}, vol.~1, no.~2, pp. 251--263,
  1973.

\bibitem{VeeravalliBasar_94}
V.~V. Veeravalli, T.~Ba\c{s}ar, and H.~V. Poor, ``Minimax robust decentralized
  detection,'' \emph{IEEE Transactions on Information Theory}, vol.~40, no.~1,
  pp. 35--40, January 1994.

\bibitem{PolWu17}
\BIBentryALTinterwordspacing
Y.~Polyanskiy and Y.~Wu, ``Lecture notes on information theory,'' August 2017.
  [Online]. Available:
  \url{http://people.lids.mit.edu/yp/homepage/data/itlectures_v5.pdf}
\BIBentrySTDinterwordspacing

\bibitem{Hoeffding_65}
W.~Hoeffding, ``Asymptotically optimal tests for multinomial distributions,''
  \emph{Annals of Mathematical Statistics}, vol.~36, no.~2, pp. 369--401, 1965.

\bibitem{TenSan81}
R.~R. Tenney and N.~R. Sandell, ``Detection with distributed sensors,''
  \emph{IEEE Transactions on Aerospace and Electronic Systems}, 1981.

\bibitem{GeoBer02}
C.~George and R.~L. Berger, \emph{Statistical inference}.\hskip 1em plus 0.5em
  minus 0.4em\relax Duxbury, 2002.

\bibitem{ZeitouniGutman_91}
O.~Zeitouni and M.~Gutman, ``On universal hypothesis testing via large
  deviations,'' \emph{IEEE Transactions on Information Theory}, vol.~37, no.~2,
  pp. 285--290, March 1991.

\bibitem{ZeitouniZiv_92}
O.~Zeitouni, J.~Ziv, and N.~Merhav, ``When is the generalized likelihood ratio
  test optimal?'' \emph{IEEE Transactions on Information Theory}, vol.~38,
  no.~5, pp. 1597--1602, 1992.

\bibitem{LamShoPea82}
L.~Lamport, R.~Shostak, and M.~Pease, ``The {B}yzantine generals problem,''
  \emph{AMC Transactions on Programming Languages and Systems}, vol.~4, July
  1982.

\bibitem{KaiHanBra15}
B.~Kailkhura, Y.~S. Han, S.~Brahma, and P.~K. Varshney, ``Asymptotic analysis
  of distributed {B}ayesian detection with {B}yzantine data,'' \emph{IEEE
  Signal Processing Letters}, vol.~22, 2015.

\bibitem{CheCheWan18}
W.-N. Chen, H.-C. Chen, and I.-H. Wang, ``On the fundamental limits of
  heterogeneous distributed detection: Price of anonymity,'' \emph{IEEE
  International Symposium on Information Theory (ISIT)}, June 2018.

\bibitem{CovTho06}
T.~M. Cover and J.~A. Thomas, \emph{Elements of Information Theory}.\hskip 1em
  plus 0.5em minus 0.4em\relax Wiley-Interscience, 2006, no. 0471241954.

\bibitem{Csi06}
I.~Csisz{\'a}r, ``A simple proof of {S}anov's theorem,'' \emph{Bull. Braz.
  Math. Soc. (N.S.)}, 2006.

\bibitem{Hol00}
F.~den Hollander, \emph{Large Deviations}, ser. Fields Institude
  Monographs.\hskip 1em plus 0.5em minus 0.4em\relax American Mathematical
  Society, 2000, no.~14.

\bibitem{DemZei10}
A.~Dembo and O.~Zeitouni, \emph{Large Deviations Techniques and Applications},
  ser. 38.\hskip 1em plus 0.5em minus 0.4em\relax Springer-Verlag, 2010, vol.
  Stochastic Modelling and Applied Probability.

\bibitem{RoyFit10}
H.~Royden and P.~Fitzpatrick, \emph{Real Analysis}.\hskip 1em plus 0.5em minus
  0.4em\relax Pearson, 2010.

\end{thebibliography}
